\documentclass[a4paper,11pt]{article}
\usepackage{jheppub}
\usepackage[T1]{fontenc} 
\usepackage{amssymb}
\usepackage{amsmath}
\usepackage{xspace}
\usepackage{caption}
\usepackage{appendix}
\usepackage{float}
\usepackage{hyperref}
\usepackage{diagbox}
\usepackage{soul}
\usepackage{subfig}
\usepackage{slashed}
\usepackage{amsfonts}
\usepackage{multirow}
\usepackage{mathrsfs}
\usepackage{graphicx}
\usepackage{amsmath}
\usepackage{amssymb}
\usepackage{bm}
\usepackage{bbm}
\usepackage{color}
\usepackage{tcolorbox,colortbl}
\usepackage{booktabs}
\usepackage{comment}

\newcommand{\nn}{\nonumber}
\newcommand{\bea}{\begin{align}}
	\newcommand{\eea}{\end{align}}
\newcommand{\beq}{\begin{equation}}
	\newcommand{\eeq}{\end{equation}}
\newcommand{\bqa}{\begin{eqnarray}}
	\newcommand{\eqa}{\end{eqnarray}}

\newcommand{\M}{\ensuremath \text{M}}

\newcommand{\eps}{\epsilon}

\newcommand{\als}{\alpha_s}

\allowdisplaybreaks[4]
\title{Analytic decay width of the Higgs boson to massive
bottom quarks at order $\alpha_s^3$}
\author[b,c]{Jian Wang,}
\author[d,e]{Xing Wang,}
\author[a]{Yefan Wang}

\affiliation[a]{Department of Physics and Institute of Theoretical Physics, Nanjing Normal University,
Nanjing, Jiangsu 210023, China}
\affiliation[b]{School of Physics, Shandong University, Jinan, Shandong 250100, China}
\affiliation[c]{Center for High Energy Physics, Peking University, Beijing 100871, China}
\affiliation[d]{
School of Science and Engineering, The Chinese University of Hong Kong,
Shenzhen, Guangdong, 518172,  China}
\affiliation[e]{
Physik Department, TUM School of Natural Sciences,
Technische Universität München, D - 85748 Garching, Germany}

\preprint{TUM-HEP-1530/24}
\abstract
{
The Higgs boson decay into bottom quarks is the dominant decay channel contributing to its total decay width, which can be used to measure the bottom quark Yukawa coupling and mass.
This decay width has been computed up to $\mathcal{O}(\alpha_s^4)$ for the process induced by the bottom quark Yukawa coupling, assuming massless final states,
and the corresponding corrections beyond $\mathcal{O}(\alpha_s^2)$ are found to be less than $0.2\%$.
We present an analytical result for the decay into massive bottom quarks at  $\mathcal{O}(\alpha_s^3)$ that includes the contribution from the top quark Yukawa coupling induced process.
We have made use of the optical theorem,  canonical differential equations and the regular basis in the calculation and expressed the result in terms of multiple polylogarithms and elliptic functions.
We propose a systematic and unified procedure to derive the $\eps$-factorized differential equation for the three-loop kite integral family, which includes the three-loop banana integrals as a sub-sector.
We find that the $\mathcal{O}(\alpha_s^3)$ corrections increase the decay width, relative to the result up to $\mathcal{O}(\alpha_s^2)$,  by $1\%$ due to the large logarithms $\log^i (m_H^2/m_b^2)$ with $ 1\le i \le 4 $ in the small bottom quark mass limit.
The coefficient of the double logarithm is proportional to $C_A-C_F$, which is the typical color structure in the resummation of soft quark contributions at subleading power.
}

\emailAdd{j.wang@sdu.edu.cn}
\emailAdd{xing.wang@tum.de}
\emailAdd{wangyefan@nnu.edu.cn}

\begin{document}
	\maketitle
	\flushbottom

\section{Introduction}

After the discovery of the Higgs boson, precise measurements of its couplings are of great importance.
The couplings with massive gauge bosons have been measured with an accuracy of $5\%$,
and the Yukawa couplings of the Higgs boson to the top quark, bottom quark, and tau lepton have uncertainties of around $10\%$ \cite{ATLAS:2024fkg}.
Among these couplings, the bottom Yukawa coupling $y_b$ stands out for its dominant contribution to the Higgs boson total decay width.
The expected uncertainty on the branching ratio measurements of $H\to b\bar{b}$ can reach $5\%$ at the HL-LHC \cite{ATLAS:2018jlh}.
The signal strength of $H\to b\bar{b}$ at the CEPC can be measured with a statistical uncertainty of $0.27\%$ after accumulating an integrated luminosity of $5.6 ~{\rm ab}^{-1}$ \cite{Zhu:2022lzv,CEPCPhysicsStudyGroup:2022uwl}.

These promising experiments call for precise theoretical predictions.
The next-to-leading order (NLO) QCD and electroweak (EW) corrections to the Higgs boson decay into massive bottom quarks have been computed long ago \cite{Braaten:1980yq,Sakai:1980fa,Janot:1989jf,Drees:1990dq,Dabelstein:1991ky,Kniehl:1991ze}.
The NNLO QCD corrections were obtained numerically  and analytically in refs. \cite{Bernreuther:2018ynm,Behring:2019oci,Somogyi:2020mmk} and refs. \cite{Wang:2023xud,Chetyrkin:1995pd,Harlander:1997xa,Primo:2018zby}, respectively.
When the final-state bottom quark is taken to be massless, the total and differential decay rates are known up to $\mathcal{O}(y_b^2 \alpha_s^4)$ \cite{Gorishnii:1990zu,Chetyrkin:1996sr,Baikov:2005rw,Herzog:2017dtz} and $\mathcal{O}(y_b^2 \alpha_s^3)$ \cite{Anastasiou:2011qx,DelDuca:2015zqa,Mondini:2019gid,Chen:2023fba}, respectively\footnote{The kinematical $\pi^2$ terms are resummed in ref. \cite{AlamKhan:2023dms}.}. 
It is found that the higher-order corrections beyond NNLO are less than $0.2\%$.
The mixed QCD$\times$EW corrections have been considered in \cite{Kataev:1997cq,Mihaila:2015lwa} and turn out to decrease the decay width by about $1\%$.
The matching to parton shower \cite{Bizon:2019tfo,Hu:2021rkt} and the higher-order corrections to the $H \to bbj $ \cite{Mondini:2019vub} and $H\to b\bar{b}b\bar{b}$ \cite{Gao:2019ypl} processes have also been investigated.
The process induced by the top quark Yukawa coupling $y_t$ could contribute to the decay width of $H\to b\bar{b}$ at two-loop level \cite{Primo:2018zby}, and even higher-order corrections have been obtained numerically when the top quark loop is integrated out in the large mass limit \cite{Mondini:2020uyy}\footnote{Such corrections to the Higgs decay into all hadronic states, including both bottom quarks and gluons, are available up to $\mathcal{O}(\alpha_s^4)$ \cite{Chetyrkin:1997vj,Davies:2017xsp} when the bottom quark mass is neglected except for that in the couplings. }.

In this paper we are going to provide an analytic form of the $H\to b\bar{b}$ decay width at QCD ${\rm N^3LO}$ by including  the finite bottom quark mass effect in the contribution induced by the top quark Yukawa coupling $y_t$.
This is an extension of our previous work \cite{Wang:2023xud}, in which we present a full analytic result of the QCD NNLO correction to the $H\to b\bar{b}$ decay induced by the bottom quark Yukawa coupling $y_b$.
We find that the finite bottom quark mass effect at $\mathcal{O}(\alpha_s^2)$ is below one percent  compared to the result with massless final-state bottom quarks at $\mathcal{O}(\alpha_s^2)$, since the corresponding decay rate can be Taylor expanded in $m_b^2/m_H^2\sim 5\times 10^{-4}$.
Therefore we neglect the bottom quark mass in the $\mathcal{O}(\alpha_s^3)$ corrections to the $y_b$ induced processes. 
By contrast, the contribution from the $y_t$ coupling is enhanced by logarithmic terms $\log^i(m_H^2/m_b^2) $ from $\mathcal{O}(\alpha_s^2)$ and even power-enhanced from  $\mathcal{O}(\alpha_s^3)$.
The former has aroused interest in the subleading power resummation recently and makes more significant contributions to the decay width than naive expectation with perturbative expansion in $\alpha_s$.
The latter indicates that the contribution is not proportional to the bottom Yukawa coupling at leading power.
These new features can only be unveiled by an analytical calculation.
Moreover, we have obtained analytical results in different  schemes of the mass renormalization so that we are able to estimate the uncertainties associating with the renormalization schemes.

The rest of this paper is organized as follows.
In section \ref{sec:sec2}, we introduce the calculation framework including the effective operators. 
The requisite master integrals are calculated  analytically in section \ref{sec:MIs}.
After a brief description of the renormalization in section \ref{sec:ren}, we present the full analytic result in section \ref{sec:anares}.
The asymptotic expansion is discussed in section \ref{sec:asyexp} and the numerical results are investigated in section \ref{sec:num}.
We conclude in section \ref{sec:conclusion}.
The topological diagrams of the master integrals are shown in appendix \ref{sec:topo}.
We describe a systematic and unified procedure to derive an $\epsilon$-factorized differential equation for the three-loop kite integral family that contains three-loop banana integrals as a sub-sector in appendix \ref{sec:K3}.

\section{Calculation framework}
\label{sec:sec2}

At the leading order (LO), the decay of a Higgs boson to a massive bottom quark pair is induced by the bottom Yukawa coupling.
However, the top quark Yukawa coupling starts to contribute to the decay at higher orders in $\alpha_s$.
The top quark loop can be integrated in the large top quark mass limit,
leading to an effective coupling between the Higgs boson and the gluons.
This approximation works exceedingly well with a relative error at the percent level \cite{Primo:2018zby}.
As such, the effective Lagrangian \cite{Inami:1982xt,Chetyrkin:1996wr,Chetyrkin:1996ke,Davies:2017xsp} 
that is relevant to Higgs boson decay to massive bottom quarks can be written as   
\begin{align}
\mathcal{L}_{\text{eff}} = -\frac{H}{v}\left(
C_1 \mathcal{O}_1^R
+C_2 \mathcal{O}_2^R
\right)+ \mathcal{L}_{\text{QCD}}\,,
\label{Leff}
\end{align}
where $v$ is the vacuum expectation value of the Higgs field, $\mathcal{L}_{\text{QCD}}$ is the QCD Lagrangian without top quarks and only bottom quarks are considered massive.   
The two renormalized operators are defined by
\begin{align}
\mathcal{O}_1^R =  Z_{11}\mathcal{O}_1 + Z_{12} \mathcal{O}_2,\quad
\mathcal{O}_2^R =  Z_{21} \mathcal{O}_1 + Z_{22} \mathcal{O}_2
\label{mix}
\end{align}
with 
\begin{align}
\mathcal{O}_1 = 
(G^{0}_{a,\mu\nu})^2,\quad
\mathcal{O}_2= m_b^{0}\bar{b}^{0}{b}^{0}.
\end{align}
The superscript ``$0$'' indicates the bare quantity.
The renormalization constant $Z_{21}=0$ in the $\overline{\text{MS}}$ scheme due to the Ward identity in the process $H\to gg$ via a bottom loop, and $Z_{22}=1$ because of our choice of the bare quantities in $\mathcal{O}_2$.
The renormalization constants $Z_{11}$ and $Z_{12}$ \cite{Chetyrkin:1996ke} can be derived from $Z_{\als}$ and $Z_m$ in the $\overline{\text{MS}}$ scheme,
\begin{align}
Z_{11}&= 1+ \als\frac{\partial \log Z_{\als}}{\partial \als}  
=  1 + \left(\frac{\als}{\pi}\right)\left(\frac{-11C_A+2n_f}{12\eps}\right)
\nonumber\\&\quad
+\left(\frac{\als}{\pi}\right)^2\left(
\frac{\left(-11C_A+2n_f\right)^2}{144\eps^2}-\frac{17C_A^2-5C_An_f-3C_Fn_f}{24\eps}
\right)+\mathcal{O}(\als^{3}),\nonumber\\
Z_{12} &= -4\als\frac{\partial \log Z^{\overline{\text{MS}}}_m}{\partial \als}= \left(\frac{\als}{\pi}\right)\left( \frac{3C_F}{\eps}\right)
\nonumber\\&\quad
+\left(\frac{\als}{\pi}\right)^2C_F\left(
\frac{-11C_A+2n_f}{4\eps^2}+\frac{97C_A+9C_F-10n_f}{24\eps}
\right)+\mathcal{O}(\als^{3}).
\end{align}
where $C_A$ and $C_F$ are the QCD color factors and $n_f$ denotes the number of active quark flavors. 
The Wilson coefficients of the effective operators are given by \cite{Chetyrkin:1996ke,Chetyrkin:1997un,Liu:2015fxa,Davies:2017xsp}
\begin{align}
C_1 &= -\left(\frac{\als}{\pi}\right)\frac{1}{12}- \left(\frac{\als}{\pi}\right)^2\frac{11}{48} 
\nonumber\\&\quad
- \left(\frac{\als}{\pi}\right)^3\left[\frac{2777}{3456}+\frac{19}{192}L_t-n_f\left(\frac{67}{1152}-\frac{1}{36}L_t\right)\right]+\mathcal{O}(\als^4),\nonumber\\
C_2 &= 1 + \left(\frac{\als}{\pi}\right)^2\left[
\frac{5}{18} - \frac{1}{3}L_t\right] 
\nonumber\\&
\quad+ \left(\frac{\als}{\pi}\right)^3 
\left[
-\frac{841}{1296}+\frac{5}{3}\zeta(3)-\frac{79}{36}L_t
-\frac{11}{12}L_t^2+n_f\left(
\frac{53}{216}+\frac{1}{18}L_t^2\right)
\right]+\mathcal{O}(\als^4),
\label{C1C2}
\end{align}
where $L_t$ = $\log(\mu^2/m_t^2)$ and $m_t$ is the top quark mass in the on-shell scheme. 
In the above expressions, the color factors $C_F=4/3$, $C_A=3$, $T_R=1/2$ have been substituted. 
Note that $C_1$ starts from $\mathcal{O}\left(\als\right)$ 
and that the first nontrivial QCD corrections in $C_2$ begin at $\mathcal{O}\left(\als^2\right)$.

According to the combination structure of effective operators in the squared amplitudes, the decay width of $H\rightarrow b\bar{b}$ can be decomposed into three parts, i.e.,
\begin{align}
\Gamma_{H\rightarrow b\bar{b}}=  \Gamma^{C_2C_2}_{H\rightarrow b\bar{b}} + \Gamma^{C_1C_2}_{H\rightarrow b\bar{b}} + \Gamma^{C_1C_1}_{H\rightarrow b\bar{b}}.
\end{align}
Each of them can be expanded in a series of the strong coupling $\als$, 
\begin{align}
\Gamma^{C_2C_2}_{H\rightarrow b\bar{b}} &= C_2C_2\left[\Delta^{C_2C_2}_{0,b\bar{b}}+\left(\frac{\als}{\pi}\right)\Delta^{C_2C_2}_{1,b\bar{b}}+\left(\frac{\als}{\pi}\right)^2\Delta^{C_2C_2}_{2,b\bar{b}}+\left(\frac{\als}{\pi}\right)^3\Delta^{C_2C_2}_{3,b\bar{b}}+\mathcal{O}(\als^4)\right],\nonumber\\
\Gamma^{C_1C_2}_{H\rightarrow b\bar{b}} &= C_1C_2\left[\left(\frac{\als}{\pi}\right)\Delta^{C_1C_2}_{1,b\bar{b}}+\left(\frac{\als}{\pi}\right)^2\Delta^{C_1C_2}_{2,b\bar{b}}+\mathcal{O}(\als^3)\right],\nonumber\\
\Gamma^{C_1C_1}_{H\rightarrow b\bar{b}} &= C_1C_1\left[\left(\frac{\als}{\pi}\right)\Delta^{C_1C_1}_{1,b\bar{b}}+\mathcal{O}(\als^2)\right],
\label{Delta}
\end{align}
where we have written down explicitly the terms that are needed to obtain $\Gamma_{H\rightarrow b\bar{b}}$ to $\mathcal{O}(\als^3)$. 
Note that the leading orders of $\Gamma^{C_1C_2}_{H\rightarrow b\bar{b}}$ and $\Gamma^{C_1C_1}_{H\rightarrow b\bar{b}}$ are $\mathcal{O}(\als^2)$ and $\mathcal{O}(\als^3)$, respectively.

The analytic results of $\Delta^{C_2C_2}_{i,b\bar{b}}$ have been calculated up to $\mathcal{O}(\als^4)$ and $\mathcal{O}(\als^2)$ in the massless\footnote{The bottom quark Yukawa coupling is still finite.} \cite{Chetyrkin:1996sr,Baikov:2005rw,Herzog:2017dtz} and massive \cite{Wang:2023xud} bottom quark cases, respectively. 
It is found that the massless result can be obtained from the massive one by taking the limit continuously.  
Therefore, it is reasonable to take $\Delta^{C_2C_2}_{3,b\bar{b}}$ calculated in \cite{Chetyrkin:1996sr,Herzog:2017dtz} as the result for the decay into massive bottom quarks, since the relative error is expected to be below one percent.

In this paper, we present analytic results of $\Delta^{C_1C_1}_{1,b\bar{b}}$, $\Delta^{C_1C_2}_{1,b\bar{b}}$ and $\Delta^{C_1C_2}_{2,b\bar{b}}$ with full dependence on the bottom quark mass.
These results cannot be Taylor expanded in $m_b^2$ because of the logarithmic dependence. 
These large logarithms make a significant contribution to the corrections and may sabotage a perturbative expansion in $\als$. 
They are different from the traditional large logarithms induced by soft gluons, 
and their all-order structure is an interesting topic under active research.

\begin{figure}[ht]
	\centering
\begin{minipage}{0.3\linewidth}
	\centering
	\includegraphics[width=0.8\linewidth]{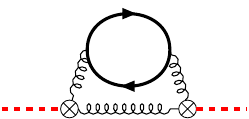}
 \caption*{(a)}
\end{minipage}
 \begin{minipage}{0.3\linewidth}
		\centering
		\includegraphics[width=0.8\linewidth]{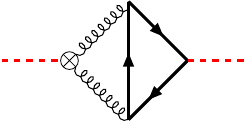}
   \caption*{(b)}
\end{minipage}
\begin{minipage}{0.3\linewidth}
		\centering
		\includegraphics[width=0.8\linewidth]{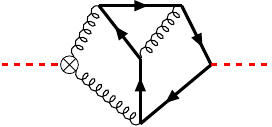}
  \caption*{(c)} 
\end{minipage}
\\~\\
\begin{minipage}{0.3\linewidth}
		\centering
		\includegraphics[width=0.8\linewidth]{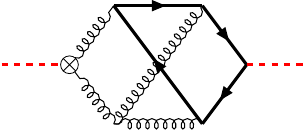}
  \caption*{(d)} 
\end{minipage}
\begin{minipage}{0.3\linewidth}
		\centering
		\includegraphics[width=0.8\linewidth]{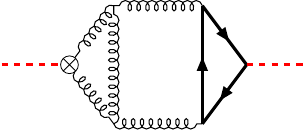}
  \caption*{(e)} 
\end{minipage}
\begin{minipage}{0.3\linewidth}
		\centering
		\includegraphics[width=0.8\linewidth]{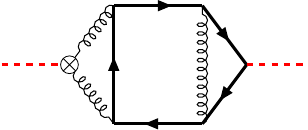}
  \caption*{(f)} 
\end{minipage}
\caption{Sample Feynman diagrams contributing to $\Delta^{C_1C_1}_{1,b\bar{b}}$, $\Delta^{C_1C_2}_{1,b\bar{b}}$ and $\Delta^{C_1C_2}_{2,b\bar{b}}$. (a) and (b) are the two-loop diagrams of
$\Delta^{C_1C_1}_{1,b\bar{b}}$ and $\Delta^{C_1C_2}_{1,b\bar{b}}$, respectively. (c)-(f) are typical three-loop diagrams of $\Delta^{C_1C_2}_{2,b\bar{b}}$. The crossed dots stand for the effective $Hgg$ vertices. The thick black and red lines
denote the massive bottom quark and the Higgs boson, respectively.}
\label{TwoThreeLoop}
\end{figure}

We compute the decay width of $H\rightarrow b\bar{b}$ via the optical theorem, i.e.,
\begin{align}
\Gamma_{H\rightarrow b\bar{b}}  = \frac{\text{Im}_{b\bar{b}}\left(\Sigma\right)}{m_H}  ,
\end{align}
where $\Sigma$ represents the amplitude of the process $H \rightarrow b \bar{b} \rightarrow H$ and the imaginary part is obtained by taking the cut on at least one bottom quark pair. 
Specifically, $\Delta^{C_1C_1}_{1,b\bar{b}}$ and $\Delta^{C_1C_2}_{1,b\bar{b}}$ require the calculation of two-loop Feynman diagrams; see, e.g.,  figure \ref{TwoThreeLoop}  (a) and (b).
$\Delta^{C_1C_2}_{2,b\bar{b}}$ is obtained after calculating the three-loop Feynman diagrams, such as figure \ref{TwoThreeLoop} (c)-(f).
Note that the imaginary part receives a contribution also from the cut on the four bottom quarks, e.g., in figure \ref{TwoThreeLoop} (c).
For clarity, we define
\begin{align}
\Delta^{C_1C_2}_{2,b\bar{b}} \equiv \tilde{\Delta}^{C_1C_2}_{2,b\bar{b}} +{\Delta}^{C_1C_2}_{2,b\bar{b}b\bar{b}}
\label{bbbb}
\end{align}
where $\tilde{\Delta}^{C_1C_2}_{2,b\bar{b}}$ represents the correction from the final states of $b\bar{b}$, $b\bar{b}g$, $b\bar{b}gg$, and $b\bar{b}q\bar{q}$,
while ${\Delta}^{C_1C_2}_{2,b\bar{b}b\bar{b}}$ gets contribution from the final state of $b\bar{b}b\bar{b}$. 

\begin{figure}[ht]
	\centering
\begin{minipage}{0.3\linewidth}
	\centering
	\includegraphics[width=0.8\linewidth]{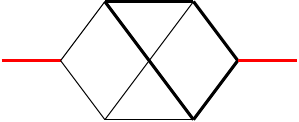}
 \caption*{NP1}
\end{minipage}
 \begin{minipage}{0.3\linewidth}
		\centering
		\includegraphics[width=0.8\linewidth]{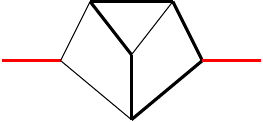}
   \caption*{P1}
\end{minipage}
\caption{The topologies of the NP1 and P1 integral families. The thick black and red lines stand for the
massive bottom quark and the Higgs boson, respectively. The other lines represent massless particles.}
\label{NP1P1}
\end{figure}

Below we focus on the three-loop calculations of $\Delta^{C_1C_2}_{2,b\bar{b}}$, as the two-loop Feynman diagrams are easier to compute. 
We have generated the Feynman diagrams and corresponding amplitudes of $H \rightarrow b \bar{b} \rightarrow H$ with effective vertices using the package {\tt FeynArts} \cite{Hahn:2000kx}. 
The model including the effective operators is implemented via {\tt FeynRules} \cite{Alloul:2013bka}. 
Then the package {\tt FeynCalc} \cite{Shtabovenko:2020gxv,Shtabovenko:2023idz} is used to simplified the Dirac matrices. 
The amplitudes are expressed as linear combinations of scalar integrals, 
which can be reduced to a set of basis integrals called master integrals (MIs) due to the identities from integration by parts (IBP) \cite{Tkachov:1981wb,Chetyrkin:1981qh}. 
In this procedure, we used the package {\tt Kira} \cite{Klappert:2020nbg} to perform the reduction. 
Some MIs can be factorized into one- and two-loop integrals and are easy to compute.
After considering the symmetries among integrals with the help of the package {\tt CalcLoop}\footnote{\url{https://gitlab.com/multiloop-pku/calcloop}}, 
the other MIs can be classified into two integral families (denoted as NP1 and P1), 
as shown in figure \ref{NP1P1}.

\section{Calculation of master integrals}
\label{sec:MIs}

\begin{figure}[ht]
	\centering
\begin{minipage}{0.3\linewidth}
	\centering
	\includegraphics[width=0.7\linewidth]{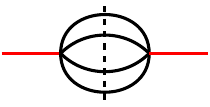}
 \caption*{(a)}
\end{minipage}
 \begin{minipage}{0.3\linewidth}
		\centering
		\includegraphics[width=0.8\linewidth]{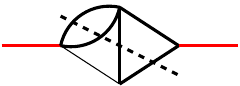}
   \caption*{(b)}
\end{minipage}
 \begin{minipage}{0.3\linewidth}
		\centering
		\includegraphics[width=0.8\linewidth]{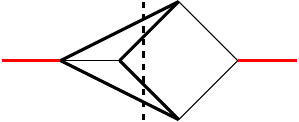}
   \caption*{(c)}
\end{minipage}
\caption{Typical MIs in calculation of ${\Delta}^{C_1C_2}_{2,b\bar{b}b\bar{b}}$. The cutting line passes through four bottom quarks. The thick black and red lines stand for the massive bottom quark and the Higgs boson, respectively. The other lines represent massless particles. }
\label{4bcut}
\end{figure}

As mentioned in eq. (\ref{bbbb}), we need to calculate the MIs with two and four bottom quark final states separately. 
In the latter case, we find 10 MIs, some of which are depicted in figure \ref{4bcut}.
One can choose the regular basis in which all MIs start from $\mathcal{O}(\eps^{0})$ and the higher order terms in $\eps$ are not required in the amplitude \cite{Lee:2019wwn}. 
The analytical results of the first 9 basis integrals can be found in \cite{Lee:2019wwn}.
They are expressed either as complete elliptic integrals of the first kind or one-fold integrals of them. 
The last MI corresponding to figure \ref{4bcut}$(c)$ appears in the basis of the integral family but does not contribute to ${\Delta}^{C_1C_2}_{2,b\bar{b}b\bar{b}}$.
However, given that this integral is interesting for the calculation of general elliptic integral families, 
we describe the procedure to construct the $\epsilon$-factorized form of the differential equation in appendix \ref{sec:K3}.

Then we turn to the calculations of MIs contributing to $\tilde{\Delta}^{C_1C_2}_{2,b\bar{b}}$. 
The topology diagrams of the MIs in the NP1 and P1 families 
are displayed in figure \ref{NP1_Topo} and figure \ref{P1_Topo}, respectively, in appendix \ref{sec:topo}.
We adopt the method of differential equations \cite{Kotikov:1991pm,Henn:2013pwa}. After choosing the proper basis, the differential equation exhibits a canonical form, i.e., the dimension parameter $\epsilon$ is factorized from the kinematic dependence. Then the differential equations can be solved recursively in terms of multiple polylogarithms (MPLs)  \cite{Goncharov:1998kja}, which are defined by $G(x)\equiv 1$ and
\bqa
	G(l_1,l_2,\ldots,l_n,x) &\equiv & \int_0^x \frac{\text{d} t}{t - l_1} G(l_2,\ldots,l_n,t)\, ,\\
	G(\overrightarrow{0}_n , x) & \equiv & \frac{1}{n!}\ln^n x\, .
\eqa
The number of elements in the set $\{l_1,l_2,\ldots,l_n\}$ is referred to as the transcendental $weight$ of the MPLs. 
Below we only show the canonical basis of the NP1 family because 
the MIs in the P1 family can be either found in the NP1 family or our previous papers \cite{Wang:2023xud,Chen:2024amk}.

\subsection{Canonical basis of the NP1 family}

The integrals in the NP1 family are represented by
\begin{align}
	I^{\rm NP1}_{n_1,n_2,\ldots,n_{9}}=\textrm{Im}_{b\bar{b}} \int \prod_{i=1}^3  \frac{(m_b^2)^{\eps}d^D q_i}{i\pi^{D/2}\Gamma(1+\eps)}~\frac{D_8^{-n_8}D_9^{-n_9}}{D_1^{n_1}~D_2^{n_2}~D_3^{n_3}~D_4^{n_4}~D_5^{n_5}~D_6^{n_6}~D_7^{n_7}}, 
\end{align}
with $D=4-2\eps$ and
all $n_i\geqslant 0,i=1,\cdots, 8$,  $n_9\leqslant 0$. 
The denominators $D_i$ read
\begin{align}
	D_1 &= q_1^2,&
	D_2 &= (q_1-q_2)^2-m_b^2,&
	D_3 &= q_2^2-m_b^2,\nonumber\\
	D_4 &= q_3^2-m_b^2,&
	D_5 &= (q_2+q_3)^2,&
	D_6 &= (q_3-k)^2-m_b^2,\nonumber\\
	D_7 &= (q_1+k)^2,&
	D_8 &=(q_1-q_2-q_3+k)^2,&
	D_9 &=(q_2-k)^2,
\end{align}
where the Feynman prescription $+i\varepsilon$ has been suppressed. 
The external momentum $k$ satisfies
$k^2 = m_H^2$. 
In the above definition of integrals, we have required only the imaginary contribution induced by a cut on a bottom quark pair. 

There are 30 master integrals contributing to $\tilde{\Delta}^{C_1C_2}_{2,b\bar{b}}$. 
To construct the canonical basis, we first select  
\begin{align}
\M^{\text{NP1}}_{1}& = \epsilon ^3 m_b^2 I^{\text{NP1}}_{0,2,2,2,0,1,0,0,0},\quad&
\M^{\text{NP1}}_{2}& = \epsilon ^3 m_b^2 I^{\text{NP1}}_{0,2,2,0,1,2,0,0,0},\quad\nonumber\\
\M^{\text{NP1}}_{3}& = \epsilon ^3 m_b^2 I^{\text{NP1}}_{0,2,2,0,2,1,0,0,0},\quad&
\M^{\text{NP1}}_{4}& = \epsilon ^3 m_b^2 I^{\text{NP1}}_{2,2,0,0,1,2,0,0,0},\quad\nonumber\\
\M^{\text{NP1}}_{5}& = \epsilon ^3 m_b^2 I^{\text{NP1}}_{2,2,0,0,2,1,0,0,0},\quad&
\M^{\text{NP1}}_{6}& = (1-2 \epsilon ) \epsilon  m_b^4 I^{\text{NP1}}_{1,3,0,0,1,3,0,0,0},\quad\nonumber\\
\M^{\text{NP1}}_{7}& = \epsilon ^3 m_b^4 I^{\text{NP1}}_{2,2,0,2,0,1,1,0,0},\quad&
\M^{\text{NP1}}_{8}& = \epsilon ^3 m_b^4 I^{\text{NP1}}_{2,0,2,0,1,2,1,0,0},\quad\nonumber\\
\M^{\text{NP1}}_{9}& = \epsilon ^3 m_b^4 I^{\text{NP1}}_{2,0,2,0,2,1,1,0,0},\quad&
\M^{\text{NP1}}_{10}& = \epsilon ^3 m_b^4 I^{\text{NP1}}_{0,2,2,2,0,1,1,0,0},\quad\nonumber\\
\M^{\text{NP1}}_{11}& = \epsilon ^3 m_b^4 I^{\text{NP1}}_{0,2,1,2,0,1,2,0,0},\quad&
\M^{\text{NP1}}_{12}& = (1-2 \epsilon ) \epsilon ^3 m_b^2 I^{\text{NP1}}_{2,2,0,1,1,1,0,0,0},\quad\nonumber\\
\M^{\text{NP1}}_{13}& = \epsilon ^3 m_b^2 I^{\text{NP1}}_{0,2,1,0,1,2,1,0,0},\quad&
\M^{\text{NP1}}_{14}& = \epsilon ^2 m_b^4 I^{\text{NP1}}_{0,3,1,0,1,2,1,0,0},\quad\nonumber\\
\M^{\text{NP1}}_{15}& = (1-2 \epsilon ) \epsilon ^4 m_b^2 I^{\text{NP1}}_{0,1,2,0,1,1,1,1,0},\quad&
\M^{\text{NP1}}_{16}& = (1-2 \epsilon ) \epsilon ^3 m_b^2 I^{\text{NP1}}_{1,2,0,1,1,0,0,2,0},\quad\nonumber\\
\M^{\text{NP1}}_{17}& = (1-2 \epsilon ) \epsilon ^4 m_b^2 I^{\text{NP1}}_{1,2,0,1,1,0,1,1,0},\quad&
\M^{\text{NP1}}_{18}& = \epsilon ^3 m_b^4 I^{\text{NP1}}_{2,0,0,2,2,1,0,1,0},\quad\nonumber\\
\M^{\text{NP1}}_{19}& = (1-2 \epsilon ) \epsilon ^4 m_b^2 I^{\text{NP1}}_{1,2,0,1,1,1,1,0,0},\quad&
\M^{\text{NP1}}_{20}& = (1-2 \epsilon ) \epsilon ^3 m_b^4 I^{\text{NP1}}_{1,2,0,2,1,1,1,0,0},\quad\nonumber\\
\M^{\text{NP1}}_{21}& = (1-2 \epsilon ) \epsilon ^3 m_b^4 I^{\text{NP1}}_{1,3,0,1,1,1,1,0,0},\quad&
\M^{\text{NP1}}_{22}& = (1-2 \epsilon ) \epsilon ^4 m_b^2 I^{\text{NP1}}_{1,1,1,0,1,2,1,0,0},\quad\nonumber\\
\M^{\text{NP1}}_{23}& = (1-2 \epsilon ) \epsilon ^4 m_b^2 I^{\text{NP1}}_{0,2,1,1,1,1,1,0,0},\quad&
\M^{\text{NP1}}_{24}& = (1-2 \epsilon ) \epsilon ^4 m_b^2 I^{\text{NP1}}_{1,1,1,1,1,0,0,2,0},\quad\nonumber\\
\M^{\text{NP1}}_{25}& = (1-2 \epsilon ) \epsilon ^4 m_b^2 I^{\text{NP1}}_{2,1,0,1,1,1,0,1,0},\quad&
\M^{\text{NP1}}_{26}& = (1-2 \epsilon ) \epsilon ^5 m_b^2 I^{\text{NP1}}_{1,1,1,1,1,1,1,0,0},\quad\nonumber\\
\M^{\text{NP1}}_{27}& = (1-2 \epsilon ) \epsilon ^4 m_b^4 I^{\text{NP1}}_{1,1,1,2,1,1,1,0,0},\quad&
\M^{\text{NP1}}_{28}& = (1-2 \epsilon ) \epsilon ^5 m_b^2 I^{\text{NP1}}_{1,1,1,1,1,1,0,1,0},\quad\nonumber\\
\M^{\text{NP1}}_{29}& = (1-2 \epsilon ) \epsilon ^4 m_b^4 I^{\text{NP1}}_{1,1,1,2,1,1,0,1,0},\quad&
\M^{\text{NP1}}_{30}& = (1-2 \epsilon ) \epsilon ^5 m_b^4 I^{\text{NP1}}_{1,1,1,1,1,1,1,1,0}.
\end{align}
Here all $\M^{\text{NP1}}_{i}$ are dimensionless. 
We define the dimensionless variable $z \equiv m_H^2/m_b^2$. Then we obtain the following canonical basis of the NP1 family: 
\begin{align}
F^{\text{NP1}}_1&=\M^{\text{NP1}}_1 r,\quad
F^{\text{NP1}}_2=\M^{\text{NP1}}_2 (-z),\quad
F^{\text{NP1}}_3=\frac{\M^{\text{NP1}}_2 r}{2}+\M^{\text{NP1}}_3 r,\quad
\nonumber\\
F^{\text{NP1}}_4&=\M^{\text{NP1}}_4 (-z),\quad
F^{\text{NP1}}_5=\left(\M^{\text{NP1}}_4+\M^{\text{NP1}}_5\right) (-r),\quad
F^{\text{NP1}}_6=\frac{\M^{\text{NP1}}_5 (z-4)}{4}+\M^{\text{NP1}}_6,\quad
\nonumber\\
F^{\text{NP1}}_7&=\M^{\text{NP1}}_7r (-z),\quad
F^{\text{NP1}}_8=\M^{\text{NP1}}_8 z^2,\quad
F^{\text{NP1}}_9=\frac{\M^{\text{NP1}}_8 r z}{2}+\M^{\text{NP1}}_9 r z,\quad
\nonumber\\
F^{\text{NP1}}_{10}&=\M^{\text{NP1}}_{10} r z,\quad
F^{\text{NP1}}_{11}=\frac{ \M^{\text{NP1}}_{10} (z-4)}{2} z+\M^{\text{NP1}}_{11} (z-4) z,\quad
\nonumber\\
F^{\text{NP1}}_{12}&=-\frac{\M^{\text{NP1}}_4 r (7 z-12)}{12 (4-z)}+\frac{\M^{\text{NP1}}_{12} r}{4-z}-\frac{\M^{\text{NP1}}_6 r}{3 (4-z)}+\frac{\M^{\text{NP1}}_5 r}{12},\quad
\nonumber\\
F^{\text{NP1}}_{13}&=\M^{\text{NP1}}_2+\M^{\text{NP1}}_4+\M^{\text{NP1}}_{13} (1-z) (2 \epsilon +1)+\M^{\text{NP1}}_{14} (4-4 z) \epsilon +\M^{\text{NP1}}_{15} \left(1-\frac{z}{4}\right),\quad
\nonumber\\
F^{\text{NP1}}_{14}&=\frac{\M^{\text{NP1}}_2 r (z-8)}{8 (z-4)}+\frac{\M^{\text{NP1}}_4 r (z-3)}{2 (z-4)}+\frac{\M^{\text{NP1}}_{13} r (2 (z+1) \epsilon +z-3)}{2 (z-4)}
\nonumber\\&\quad
+\frac{\M^{\text{NP1}}_{14} r (2 z \epsilon -1)}{z-4}
+\frac{\M^{\text{NP1}}_{15} r (z-3)}{2 (z-4)},\quad
F^{\text{NP1}}_{15}=\M^{\text{NP1}}_{15} (-z),\quad
\nonumber\\
F^{\text{NP1}}_{16}&=\frac{\M^{\text{NP1}}_4 r}{3}-\frac{\M^{\text{NP1}}_5 r (z-4)}{6 z}-\frac{2 \M^{\text{NP1}}_6 r}{3 z}-\frac{\M^{\text{NP1}}_8 r z}{2}+\M^{\text{NP1}}_{16} r+\M^{\text{NP1}}_{17} r,\quad
\nonumber\\
F^{\text{NP1}}_{17}&=\M^{\text{NP1}}_{17} (-z),\quad
F^{\text{NP1}}_{18}=\M^{\text{NP1}}_{18} (-r) z,\quad
F^{\text{NP1}}_{19}=\M^{\text{NP1}}_{19} (-z),\quad
\nonumber\\
F^{\text{NP1}}_{20}&=\frac{\M^{\text{NP1}}_4 r}{3}-\frac{\M^{\text{NP1}}_5 r (z-4)}{6 z}-\frac{2 \M^{\text{NP1}}_6 r}{3 z}+\frac{3 \M^{\text{NP1}}_{19} r}{2}+\M^{\text{NP1}}_{20} r+\M^{\text{NP1}}_{21} r,\quad
\nonumber\\
F^{\text{NP1}}_{21}&=\M^{\text{NP1}}_{21} (-z),\quad
F^{\text{NP1}}_{22}=\M^{\text{NP1}}_{22} (-z),\quad
F^{\text{NP1}}_{23}=\M^{\text{NP1}}_{23} (-z),\quad
F^{\text{NP1}}_{24}=\M^{\text{NP1}}_{24} (-z),\quad
\nonumber\\
F^{\text{NP1}}_{25}&=\M^{\text{NP1}}_{25} (-z),\quad
F^{\text{NP1}}_{26}=\M^{\text{NP1}}_{26} (-z),\quad
\nonumber\\
F^{\text{NP1}}_{27}&=\frac{\M^{\text{NP1}}_2 (z+4)r}{4 (z-2)}+\frac{\M^{\text{NP1}}_4 (2z+3)r}{3 (z-2)}-\frac{\M^{\text{NP1}}_5 (z-4)r}{12 (z-2)}+\frac{\M^{\text{NP1}}_6 r}{6-3 z}+\frac{\M^{\text{NP1}}_8 r z^2}{8-4 z}
\nonumber\\ &\quad
+\frac{\M^{\text{NP1}}_{13} (2 \epsilon +1) (1-z)r}{z-2}+\frac{\M^{\text{NP1}}_{15} (z+2)r}{2 (z-2)}+\frac{4 \M^{\text{NP1}}_{14} \epsilon  (1-z)r}{z-2}+\frac{\M^{\text{NP1}}_{19} r z}{2 (z-2)}
\nonumber\\& \quad
+\frac{\M^{\text{NP1}}_{21} r z}{z-2}+\frac{\M^{\text{NP1}}_{22} r z}{4-2 z}+\frac{\M^{\text{NP1}}_{26} r z}{2-z}+\frac{\M^{\text{NP1}}_{27} r z}{2-z},\quad
F^{\text{NP1}}_{28}=\M^{\text{NP1}}_{28} (-z),\quad
\nonumber\\
F^{\text{NP1}}_{29}&=\frac{\M^{\text{NP1}}_{24} r}{2}-\frac{\M^{\text{NP1}}_{25} r}{2}+\M^{\text{NP1}}_{28} r+\M^{\text{NP1}}_{29} r,\quad
F^{\text{NP1}}_{30}=\M^{\text{NP1}}_{30} z^2\,,
\end{align}
where
\begin{align}
\label{eq:rationalization}
r = \sqrt{z(z-4)}.
\end{align}
To rationalize the square root $r$, we write
\begin{align}
    \label{eq:rationalizationinv}
	z = -\frac{(w-1)^2}{w}
\end{align}
with  $-1<w<0$.
Then $r$ is rationalized to 
\begin{align}
r = \frac{(w+1) (w-1)}{w}.
\end{align}

The differential equations for the canonical basis $\boldsymbol{F}^{\text{NP1}}=(F^{\text{NP1}}_{1},\cdots,F^{\text{NP1}}_{30})^T$ become 
\begin{align}
\frac{d \boldsymbol{F}^{\text{NP1}}}{dw} = \epsilon \left( \sum_{i=1}^5 \frac{\boldsymbol{N_i}^{\text{NP1}}}{w-l_i^{\text{NP1}}}\right) \boldsymbol{F}^{\text{NP1}}
\end{align}
with $\boldsymbol{N_i}^{\text{NP1}}$ being constant rational matrices and
\begin{align}
l^{\text{NP1}}_1 = 0\,,~ 
l^{\text{NP1}}_{2} = 1\,,~
l^{\text{NP1}}_{3} = -1\,,~
l^{\text{NP1}}_{4} = \frac{1+\sqrt{3} i}{2}\,,~
l^{\text{NP1}}_{5} = \frac{1-\sqrt{3} i}{2}.
\end{align}

\subsection{Boundary conditions}
After solving the canonical differential equations, the results are written as the linear combinations of MPLs \cite{Goncharov:1998kja} with undetermined integration constants,
where the MPLs can be evaluated using the {\tt PolyLogTools} package \cite{Duhr:2019tlz} based on {\tt GiNac} \cite{Bauer:2000cp}. 
In another method, 
we get the high-precision numerical results of the basis integrals using the package {\tt AMFlow} \cite{Liu:2017jxz,Liu:2020kpc,Liu:2022chg,Liu:2022mfb}. 
Notice that we have set the specific cuts corresponding to a bottom quark pair, excluding the cuts on four bottom quarks or two gluons.
Explicit examples with more details on this method can be found in \cite{Chen:2024amk}. 
Comparing the two results, we obtain the numerical results of the integration constants.
The analytical form is reproduced using the PSLQ algorithm \cite{Ferguson:1999aa}. 
It turns out that the constants consist of $\zeta(n)$ and $\pi^n$. 

\section{Renormalization}
\label{sec:ren}

The results of the cut three-loop diagrams contain ultra-violet divergences, which would cancel after considering the renormalization of the mass, couplings, and effective operators.

In our calculation, the Higgs boson and the bottom quark are taken to be massive while the other light quarks are massless. The bottom quark mass appears in the Yukawa coupling and the loop propagators. 
In the Yukawa coupling, the bottom quark mass is renormalized in the $\overline{\text{MS}}$ scheme as in \cite{Wang:2023xud}, i.e., $m_b^0 = Z_{m_b}^{\overline{\rm MS}}\overline{m_b}$, in order to avoid large logarithms in $\Gamma^{C_2C_2}_{b\bar{b}}$; see e.g., \cite{Bernreuther:2018ynm, Wang:2023xud} for more details. 
The bottom quark mass in the propagators is renormalized in the on-shell scheme by default, i.e., $m_b^0= Z_m^{\rm OS} m_b$,
and a comparison of the results in the $\overline{\text{MS}}$ and on-shell schemes is given in section \ref{sec:num}.
The renormalization of the strong coupling $\als$ is performed in the $\overline{\text{MS}}$ scheme, i.e., $\als^{0}C^{\eps} = \mu^{2\eps} \als Z_{\als}$ with $ C_{\eps} = (4\pi e^{- \gamma_E} )^{\eps} $.
The needed renormalization constants are given by \cite{Caswell:1974gg,Jones:1974mm,Czakon:2007wk,Melnikov:2000qh}
\begin{align}
Z_{\als} &= 1+ \left(\frac{\als}{4\pi}\right)\frac{2n_l-31}{3\eps} +  \left(\frac{\als}{4\pi}\right)^2\left(\frac{\left(31-2 n_l\right)^2}{9 \epsilon ^2}+\frac{19 n_l-134}{3\epsilon }\right),\nonumber\\
Z^{\overline{\rm MS}}_{m_b} &= 1 - \left(\frac{\als}{4\pi}\right)\frac{4}{\eps} +\left(\frac{\als}{4\pi}\right)^2\left(\frac{2 (43-2n_l)}{3 \eps^2}+\frac{10n_l-293}{9\eps}\right),\nonumber\\
Z^{\text{OS}}_{m_b} &= 1 +\left(\frac{\als}{4\pi}\right)D_{\eps}\left(-\frac{4}{\epsilon }-\frac{16}{3}-\frac{32 \epsilon }{3}-\frac{64 \epsilon ^2}{3}\right)\nonumber\\
&\quad+\left(\frac{\als}{4\pi}\right)^2D_{\eps}^2\bigg(\frac{86-4 n_l}{3\eps^2}+\frac{10 n_l-101}{9 \epsilon }+\frac{4 \left(2 n_l-31\right)}{3 \epsilon}\log \left(\frac{\mu^2}{m_b^2}\right)\nonumber\\
&\quad+\frac{2(31-2 n_l)}{3}\log^2\left(\frac{\mu^2}{m_b^2}\right) +\frac{16(2 n_l-31)}{9}\log\left(\frac{\mu^2}{m_b^2}\right)
+\frac{\left(10 n_l-79\right)\pi ^2}{9}\nonumber\\
&\quad
+\frac{ \left(142 n_l-2251\right)}{18}+\frac{8 \zeta (3)}{3}-\frac{16\pi ^2 \log (2)}{9}
\bigg),
\end{align}
where $n_l=n_f-1$ is the number of massless quarks and 
\begin{align}
     D_{\eps} \equiv e^{\gamma_E\eps} \Gamma(1+\eps) \left(\frac{ \mu^2}{m_b^2}\right)^{\eps} .
\end{align}

Besides, the mixing between the two effective operators indicates
\begin{align}
\langle b\bar{b} | \mathcal{O}_1^R | H\rangle 
=Z_{11} \langle b\bar{b} | \mathcal{O}_1 | H\rangle 
+Z_{12} \langle b\bar{b} | \mathcal{O}_2 | H\rangle \,.
\end{align}
Part of the ultra-violet divergences in the loop amplitude of $\langle b\bar{b} | \mathcal{O}_1 | H\rangle $ has to be canceled with $Z_{12} \langle b\bar{b} | \mathcal{O}_2 | H\rangle$.
Because $Z_{12}$ starts from $\mathcal{O}(\als)$, 
the amplitude of $\langle b\bar{b} | \mathcal{O}_2 | H\rangle$ is of lower loop order.
The bottom quark mass in $\langle b\bar{b} | \mathcal{O}_1 | H\rangle $ appears only in the propagator, 
and thus the mass in $\mathcal{O}_2$ must also be renormalized in the on-shell scheme to ensure the cancellation of divergences.
After cancellation of divergences, we are free to convert the mass from one renormalization scheme to another.

\section{Analytic results}
\label{sec:anares}

After renormalization, we obtain the finite analytic results. The result of ${\Delta}^{C_1C_1}_{1,b\bar{b}}$ is 
\begin{align}
{\Delta}^{C_1C_1}_{1,b\bar{b}} \,&= \, \frac{m_H^3}{12v^2\pi}C_AC_F
\bigg[
\frac{2 (w+1)^2 \left(w^4-8 w^3+12 w^2-8 w+1\right)}{(w-1)^6}
\nonumber\\&
\quad\times\left(-G(0,w)+i \pi \right)
+\frac{(w+1)^3 \left(7 w^2-16 w+7\right)}{(w-1)^5}
\bigg].  
\end{align}
And  ${\Delta}^{C_1C_2}_{1,b\bar{b}}$ reads 
\begin{align}
&{\Delta}^{C_1C_2}_{1,b\bar{b}} \,= \, \frac{m_Hm_b\overline{m_b}(\mu)}{v^2\pi}C_AC_F\bigg[
\frac{i w \left(3 w^3+w^2+8 w-5\right) (\pi +i G(0,w))}{2 (w-1)^4}
\nonumber\\&
+\frac{(w+1)^2 \left(i \pi  G(0,w)-4 i \pi  G(1,w)-G(0,0,w)-2 G(0,1,w)+4 G(1,0,w)+\pi ^2\right)}{4 (w-1)^2}
\nonumber\\&
+\frac{3 (w+1)^3}{4 (w-1)^3}\left(2 G(1,w)+\log \left(\frac{\mu ^2}{m_H^2}\right)\right)
+\frac{19 w^3+27 w^2+27 w+19}{8 (w-1)^3}
\bigg],
\end{align}
where the factor $\overline{m_b}$ comes from the Yukawa coupling while $m_b$ arises from the propagator due to the requirement of a helicity flip along the fermion line coupled to a Higgs boson. 

The higher-order result, $\Delta^{C_1C_2}_{2,b\bar{b}}$, includes contributions from both the two and four bottom quark final states, as shown in eq. (\ref{bbbb}).
The four bottom quark final state contribution is  given by
\begin{align}
&{\Delta}^{C_1C_2}_{2,b\bar{b}b\bar{b}} \,=\,\frac{m_Hm_b\overline{m_b}(\mu)}{v^2}
\frac{1}{6912 \pi^2  z^2}\bigg[
-3 \left(279 z^2-1284 z+1024\right)F^{4b}_1(z)
\nonumber\\&
+6\left(77 z^2-764 z+1376\right)F^{4b}_2(z) 
+24(z-4) (71 z-172)F^{4b}_3(z)
\nonumber\\&
+16 \left(49 z^2-266 z+256\right)\beta F^{4b}_4(z)
-24z^2F^{4b}_5(z)+48 (5 z-18) z \beta F^{4b}_6(z)
\nonumber\\&
-192\left(z^2-3 z-7\right)F^{4b}_7(z)
-48(5 z-18) z \beta F^{4b}_8(z)-96\left(z^2-4 z+2\right)F^{4b}_9(z)
\bigg]\,,
\end{align}
where $\beta = \sqrt{1-4/z}$ measures the velocity of the bottom quark and all the color factors $C_F=4/3$, $C_A=3$ have been substituted. The complete expressions of all $F^{4b}_i(z)$ can be found in \cite{Lee:2019wwn}. Their asymptotic expressions in the limit of $z\rightarrow \infty$ are collected in the appendix of ref. \cite{Wang:2023xud}. 

The contribution from the two bottom quark final state is 
\begin{align}
&\quad \tilde{\Delta}^{C_1C_2}_{2,b\bar{b}} \,=\, \frac{m_Hm_b\overline{m_b}(\mu)}{v^2\pi}\frac{(-w)}{(w-1)^2}\times
\nonumber\\
&\bigg(\bigg[\frac{\pi ^2 \left(101 w^3+99 w^2+93 w+91\right)}{6 (w-1) w}-\frac{i \pi  (w+1) \left(5 w^2-2 w+101\right)}{3 (w-1) w}-\frac{\left(11 w^2-8 w+11\right)n_l}{3 w}
\nonumber\\&
+\frac{759 w^6-1586 w^5+3148 w^4-3200 w^3+3148 w^2-1586 w+759}{6 w \left(w^2-w+1\right)^2}\bigg]
\frac{\left(w-1\right)G(-1,w)}{w+1}
\nonumber\\&
+\bigg[\frac{\left(17 w^4-15\right) \zeta (3)}{3 (w-1)}
+\frac{i \pi ^3 \left(19 w^4-108 w^3-144 w^2-108 w+19\right)}{18 (w-1)}
\nonumber\\&
+\frac{\pi ^2n_l(w-1) (w+1)^2}{3}
+\frac{\pi ^2 \left(379 w^4+236 w^3-444 w^2-84 w+53\right)}{36 (w-1)}
\nonumber\\&
-\frac{i \pi}{72 (w-1)^2 \left(w^2-w+1\right)^3}
(8463 w^{11}-33072 w^{10}+69555 w^9-96715 w^8
\nonumber\\&
+98026 w^7
-87511 w^6
+76159 w^5-69754 w^4+52531 w^3-29563 w^2+9664 w-1303)
\nonumber\\&
+\frac{i \pi  \left(23 w^3+53 w^2+w-5\right) n_l}{9}
-\frac{1}{144 (w-1) \left(w^2-w+1\right)^2}(2217 w^8-15298 w^7
\nonumber\\&
-2595 w^6
+41994 w^5-102842 w^4+102798 w^3-66873 w^2+18722 w-681)
\nonumber\\&
-\frac{\left(27 w^4+64 w^3+272 w^2-296 w-21\right) n_l}{36 (w-1)}
\bigg]\frac{G(0,w)}{(w-1)w}
\nonumber\\&
+\bigg[
-\frac{\pi ^2 \left(323 w^4+48 w^3-1336 w^2-192 w+5\right)}{18 (w-1)^2}
-\frac{2(w+1)^2\pi ^2  n_l}{9}
\nonumber\\&
+\frac{i \pi}{9 (w-1)^2 \left(w^2-w+1\right)^3}
\times(572 w^{10}-2307 w^9+1796 w^8+2473 w^7-11536 w^6
\nonumber\\&
+15130 w^5
-12130 w^4+3067 w^3
+1796 w^2-2505 w+770)
\nonumber\\&
-\frac{i \pi  \left(11 w^4-214 w^2+72 w+47\right) n_l}{9 (w-1)^2}
+\frac{\left(15 w^2+22 w+15\right) (w+1) n_l}{3 (w-1)}
\nonumber\\&
-\frac{\left(2553 w^6-674 w^5+1396 w^4+3004 w^3+1396 w^2-674 w+2553\right) (w+1)}{24 (w-1) \left(w^2-w+1\right)^2}
\bigg]\frac{G(1,w)}{w}
\nonumber\\&
+\bigg[(w+1) \left(5 w^2-2 w+101\right)+i \pi  \left(101 w^3+99 w^2+93 w+91\right)\bigg]
\frac{(w+1)G(-1,0,w)}{3 (w-1)^2 w}
\nonumber\\&+
\bigg[(w-1) (w+1)-i \pi  \left(w^2+1\right)\bigg]\frac{32(w+1)^2G(-1,1,w)}{(w-1)^2 w}
\nonumber\\&
+\bigg[\frac{ \left(15-17 w^4\right)\pi ^2}{3}-\frac{2i \pi  (w-1) \left(15 w^3+31 w^2-5 w-5\right)}{3}
-\frac{1}{3 \left(w^2-w+1\right)^3}
\nonumber\\&
\times(319 w^{10}-905 w^9+2704 w^8-4881 w^7+7881 w^6-8794 w^5+8169 w^4-5169 w^3
\nonumber\\&
+2704 w^2-809 w+223)+\frac{2n_l\left(3 w^4-4 w^3+16 w^2-4 w+3\right)}{3}
 \bigg]\frac{G(0,-1,w)}{(w-1)^2 w}
\nonumber\\&
+\bigg[-\frac{\pi ^2 \left(33 w^4-162 w^3-216 w^2-162 w+25\right)}{18 (w-1)^2 w}
\nonumber\\&
+\frac{i \pi  \left(137 w^4+84 w^3-162 w^2-28 w+25\right)}{12 (w-1)^2 w}
-\frac{i \pi (w+1)^2n_l }{3 w}-\frac{n_l\left(23 w^3+53 w^2+w-5\right)}{9 (w-1) w}
\nonumber\\&
+\frac{1}{72 (w-1)^3 w \left(w^2-w+1\right)^3}(8463 w^{11}-33072 w^{10}+69555 w^9-96715 w^8+98026 w^7
\nonumber\\&
-87511 w^6+76159 w^5-69754 w^4+52531 w^3-29563 w^2+9664 w-1303)G(0,0,w)
\bigg]
\nonumber\\&
+\bigg[
\frac{\pi ^2 \left(175 w^4-216 w^3-288 w^2-216 w-121\right)}{9 (w-1)^2 w}
-\frac{i \pi  \left(98 w^4+19 w^3-128 w^2-5 w+83\right)}{3 (w-1)^2 w}
\nonumber\\&
+\frac{2 i \pi (w+1)^2n_l}{3 w}-\frac{\left(37 w^3+61 w^2+47 w-1\right)n_l}{9 (w-1) w}
+\frac{1}{12 (w-1)^2 w \left(w^2-w+1\right)^3}
\nonumber\\&
\times(271 w^{10}-1053 w^9+1754 w^8-2365 w^7+2041 w^6-2702 w^5
\nonumber\\&
+2833 w^4-3157 w^3+1754 w^2-789 w+7)\bigg]G(0,1,w)
\nonumber\\&
+\bigg[12 (w-1) (w+1)^3-i \pi  \left(11 w^4+20 w^3+34 w^2+20 w+11\right)\bigg]\frac{8G(1,-1,w)}{3 (w-1)^2 w}
\nonumber\\&
+\bigg[
-3 i \pi  \left(157 w^4-28 w^3-734 w^2-156 w-7\right)+2\left(11 w^4-214 w^2+72 w+47\right)n_l
\nonumber\\&
-\frac{2}{\left(w^2-w+1\right)^3}(572 w^{10}-2307 w^9+1796 w^8+2473 w^7-11536 w^6+15130 w^5
\nonumber\\&
-12130 w^4+3067 w^3+1796 w^2-2505 w+770)
\bigg]\frac{G(1,0,w)}{18 (w-1)^2 w}
\nonumber\\&
+\bigg[\frac{3 i \pi  \left(19 w^4-60 w^3-174 w^2-60 w+19\right)}{2 (w-1)}-4 i \pi (w-1) (w+1)^2n_l
\nonumber\\&
-81 (w+1)^3+6n_l(w+1)^3\bigg]\frac{2G(1,1,w)}{3 (w-1) w}
\nonumber\\&
-\frac{(w+1)(101 w^3+99 w^2+93 w+91)G(-1,0,0,w)}{3 (w-1)^2 w}
\nonumber\\&
+\bigg[(w-1) \left(15 w^3+31 w^2-5 w-5\right)-i \pi  \left(17 w^4-15\right)\bigg]\frac{2G(0,-1,0,w)}{3 (w-1)^2 w}
\nonumber\\&
+\bigg[2 i \pi  \left(5 w^4-3\right)-29 w^4+105 w^3+31 w^2+105 w-29
\nonumber\\&
+2(w-1)^2 (w+1)^2n_l\bigg]\frac{2G(0,0,-1,w)}{3 (w-1)^2 w}
\nonumber\\&
+\bigg[
-4 i \pi  \left(7 w^4-27 w^3-36 w^2-27 w+3\right)-137 w^4-84 w^3+162 w^2+28 w-25
\nonumber\\&
+4 (w-1)^2 (w+1)^2n_l
\bigg]\frac{G(0,0,0,w)}{12 (w-1)^2 w}
\nonumber\\&
+\bigg[
6 i \pi  \left(17 w^4-15\right)-117 w^4+146 w^3+110 w^2+242 w-15
\nonumber\\&
+4(w-1)^2 (w+1)^2n_l
\bigg]\frac{G(0,0,1,w)}{6 (w-1)^2 w}
\nonumber\\&
+\bigg[
2 i \pi  \left(37 w^4-54 w^3-72 w^2-54 w-23\right)+98 w^4+19 w^3-128 w^2+2-5w+83
\nonumber\\&
-2(w-1)^2 (w+1)^2n_l
\bigg]\frac{G(0,1,0,w)}{3 (w-1)^2 w}
\nonumber\\&
+\bigg[
32 i \pi  (w-1) \left(w^2+1\right) (w+1)-2 \left(19 w^2+58 w+19\right) (w+1)^2
\nonumber\\&
-4(w-1)^2 (w+1)^2n_l
\bigg]\frac{G(0,1,1,w)}{3 (w-1)^2 w}+\frac{8(11 w^4+20 w^3+34 w^2+20 w+11)G(1,-1,0,w)}{3 (w-1)^2 w}
\nonumber\\&
+\bigg[
-8(w-1)^2 (w+1)^2n_l-16 \left(5 w^4+15 w^3+8 w^2+15 w+5\right)
\bigg]\frac{G(1,0,-1,w)}{3 (w-1)^2 w}
\nonumber\\&
+\frac{(157 w^4-28 w^3-734 w^2-156 w-7)G(1,0,0,w)}{6 (w-1)^2 w}-\frac{2(13 w^2+30 w+13)G(1,0,1,w)}{3 w}
\nonumber\\&
+\bigg[
-3 \left(19 w^4-60 w^3-174 w^2-60 w+19\right)+8(w-1)^2 (w+1)^2n_l
\bigg]\frac{G(1,1,0,w)}{3 (w-1)^2 w}
\nonumber\\&
-\frac{4(5 w^4-3)G(0,0,-1,0,w)}{3 (w-1)^2 w}+\frac{(7 w^4-27 w^3-36 w^2-27 w+3)G(0,0,0,0,w)}{3 (w-1)^2 w}
\nonumber\\&
+\frac{(29 w^4-54 w^3-72 w^2-54 w-11)G(0,0,0,1,w)}{3 (w-1)^2 w}
\nonumber\\&
-\frac{2(37 w^4-54 w^3-72 w^2-54 w-23)G(0,1,0,0,w)}{3 (w-1)^2 w}
\nonumber\\&
+\frac{(17 w^4-15)}{3 (w-1)^2 w}\bigg[2G(0,-1,0,0,w)-3G(0,0,1,0,w)\bigg]
\nonumber\\&
+\frac{27 w^4+w^3+11 w^2+w+27}{9 (w-1)^2 w}\bigg[8\pi ^2G(x_1,w)+12i\pi G(x_1,0,w)+3i\pi G(x_1,1,w)
\nonumber\\&
+18G(x_1,0,-1,w)-12G(x_1,0,0,w)+9G(x_1,0,1,w)-3G(x_1,1,0,w)
\nonumber\\&
+8\pi ^2G(x_2,w)+12i\pi G(x_2,0,w)+3i\pi G(x_2,1,w)
+18G(x_2,0,-1,w)
\nonumber\\&
-12G(x_2,0,0,w)+9G(x_2,0,1,w)-3G(x_2,1,0,w)\bigg]
\nonumber\\&
-\frac{8 (w+1)}{3 (w-1)^2 w}\bigg[
8\pi ^2G(0,x_1,w)w^3+8\pi ^2G(0,x_2,w)w^3-12G(-1,0,1,w)w^3
\nonumber\\&
-12G(-1,1,0,w)w^3-4i\pi G(0,-1,1,w)w^3+16G(0,-1,1,w)w^3
\nonumber\\&
-4i\pi G(0,1,-1,w)w^3+16G(0,1,-1,w)w^3+12i\pi G(0,x_1,0,w)w^3
\nonumber\\&
+3i\pi G(0,x_1,1,w)w^3+12i\pi G(0,x_2,0,w)w^3+3i\pi G(0,x_2,1,w)w^3
\nonumber\\&
+4G(0,-1,0,1,w)w^3+4G(0,-1,1,0,w)w^3-8G(0,0,-1,1,w)w^3
\nonumber\\&
-4G(0,0,0,-1,w)w^3-8G(0,0,1,-1,w)w^3-6G(0,0,1,1,w)w^3
\nonumber\\&
+4G(0,1,-1,0,w)w^3-20G(0,1,0,-1,w)w^3-3G(0,1,0,1,w)w^3
\nonumber\\&
+4G(0,1,1,0,w)w^3+18G(0,x_1,0,-1,w)w^3-12G(0,x_1,0,0,w)w^3
\nonumber\\&
+9G(0,x_1,0,1,w)w^3-3G(0,x_1,1,0,w)w^3+18G(0,x_2,0,-1,w)w^3
\nonumber\\&
-12G(0,x_2,0,0,w)w^3+9G(0,x_2,0,1,w)w^3-3G(0,x_2,1,0,w)w^3
\nonumber\\&
-8\pi ^2G(0,x_1,w)w^2-8\pi ^2G(0,x_2,w)w^2-12G(-1,0,1,w)w^2
\nonumber\\&
-12G(-1,1,0,w)w^2+4i\pi G(0,-1,1,w)w^2+8G(0,-1,1,w)w^2
\nonumber\\&
+4i\pi G(0,1,-1,w)w^2+8G(0,1,-1,w)w^2-12i\pi G(0,x_1,0,w)w^2
\nonumber\\&
-3i\pi G(0,x_1,1,w)w^2-12i\pi G(0,x_2,0,w)w^2-3i\pi G(0,x_2,1,w)w^2
\nonumber\\&
-4G(0,-1,0,1,w)w^2-4G(0,-1,1,0,w)w^2+8G(0,0,-1,1,w)w^2
\nonumber\\&
+4G(0,0,0,-1,w)w^2+8G(0,0,1,-1,w)w^2+6G(0,0,1,1,w)w^2
\nonumber\\&
-4G(0,1,-1,0,w)w^2+20G(0,1,0,-1,w)w^2+3G(0,1,0,1,w)w^2
\nonumber\\&
-4G(0,1,1,0,w)w^2-18G(0,x_1,0,-1,w)w^2+12G(0,x_1,0,0,w)w^2
\nonumber\\&
-9G(0,x_1,0,1,w)w^2+3G(0,x_1,1,0,w)w^2-18G(0,x_2,0,-1,w)w^2
\nonumber\\&
+12G(0,x_2,0,0,w)w^2-9G(0,x_2,0,1,w)w^2+3G(0,x_2,1,0,w)w^2
\nonumber\\&
+8\pi ^2G(0,x_1,w)w+8\pi ^2G(0,x_2,w)w-12G(-1,0,1,w)w
\nonumber\\&
-12G(-1,1,0,w)w-4i\pi G(0,-1,1,w)w+8G(0,-1,1,w)w
\nonumber\\&
-4i\pi G(0,1,-1,w)w+8G(0,1,-1,w)w+12i\pi G(0,x_1,0,w)w
\nonumber\\&
+3i\pi G(0,x_1,1,w)w+12i\pi G(0,x_2,0,w)w+3i\pi G(0,x_2,1,w)w
\nonumber\\&
+4G(0,-1,0,1,w)w+4G(0,-1,1,0,w)w-8G(0,0,-1,1,w)w
\nonumber\\&
-4G(0,0,0,-1,w)w-8G(0,0,1,-1,w)w-6G(0,0,1,1,w)w
\nonumber\\&
+4G(0,1,-1,0,w)w-20G(0,1,0,-1,w)w-3G(0,1,0,1,w)w
\nonumber\\&
+4G(0,1,1,0,w)w+18G(0,x_1,0,-1,w)w-12G(0,x_1,0,0,w)w
\nonumber\\&
+9G(0,x_1,0,1,w)w-3G(0,x_1,1,0,w)w+18G(0,x_2,0,-1,w)w
\nonumber\\&
-12G(0,x_2,0,0,w)w+9G(0,x_2,0,1,w)w-3G(0,x_2,1,0,w)w
\nonumber\\&
-8\pi ^2G(0,x_1,w)-8\pi ^2G(0,x_2,w)-12G(-1,0,1,w)
\nonumber\\&
-12G(-1,1,0,w)+4i\pi G(0,-1,1,w)+16G(0,-1,1,w)
\nonumber\\&
+4i\pi G(0,1,-1,w)+16G(0,1,-1,w)-12i\pi G(0,x_1,0,w)
\nonumber\\&
-3i\pi G(0,x_1,1,w)-12i\pi G(0,x_2,0,w)-3i\pi G(0,x_2,1,w)
\nonumber\\&
-4G(0,-1,0,1,w)-4G(0,-1,1,0,w)+8G(0,0,-1,1,w)
\nonumber\\&
+4G(0,0,0,-1,w)+8G(0,0,1,-1,w)+6G(0,0,1,1,w)
\nonumber\\&
-4G(0,1,-1,0,w)+20G(0,1,0,-1,w)+3G(0,1,0,1,w)
\nonumber\\&
-4G(0,1,1,0,w)-18G(0,x_1,0,-1,w)+12G(0,x_1,0,0,w)
\nonumber\\&
-9G(0,x_1,0,1,w)+3G(0,x_1,1,0,w)-18G(0,x_2,0,-1,w)
\nonumber\\&
+12G(0,x_2,0,0,w)-9G(0,x_2,0,1,w)+3G(0,x_2,1,0,w)
\bigg]
\nonumber\\
&+\bigg[
\frac{\pi ^4 \left(141 w^4-1080 w^3-1440 w^2-1080 w+253\right)}{360 (w-1)^2 w}
-\frac{i \pi  \left(17 w^4-15\right) \zeta (3)}{3 (w-1)^2 w}
\nonumber\\&
-\frac{\left(62 w^4-93 w^3+42 w^2+55 w+135\right) \zeta (3)}{3 (w-1)^2 w}
+\frac{4 (w+1)^2 \zeta (3)n_l}{3 w}
\nonumber\\&
-\frac{i \pi ^3 \left(121 w^4+76 w^3-141 w^2-28 w+14\right)}{18 (w-1)^2 w}
+\frac{2 i \pi ^3(w+1)^2n_l}{9 w}
\nonumber\\&
-\frac{\pi ^2}{216 (w-1)^3 w \left(w^2-w+1\right)^3}(18781 w^{11}-75856 w^{10}+168361 w^9-254009 w^8
\nonumber\\&
+290798 w^7-291965 w^6+268037 w^5-232550 w^4+164321 w^3-88657 w^2
\nonumber\\&
+29512 w-4837)
+\frac{\pi ^2\left(43 w^4+38 w^3-58 w^2-22 w+13\right)n_l}{18 (w-1)^2 w}
\nonumber\\&
+\frac{i \pi}{144 (w-1)^2 w \left(w^2-w+1\right)^2}
(2217 w^8-15298 w^7-2595 w^6+41994 w^5-102842 w^4
\nonumber\\&
+102798 w^3-66873 w^2+18722 w-681)
+\frac{i \pi\left(27 w^4+64 w^3+272 w^2-296 w-21\right)n_l}{36 (w-1)^2 w}
\nonumber\\&
-\frac{(w+1) \left(21609 w^2+6184 w+21609\right)}{96 (w-1) w}
+\frac{(w+1) \left(308 w^2+w+308\right)n_l}{36 (w-1) w}
\bigg]\bigg)
\end{align}
with $x_1= l_4^{\rm NP1}$ and $x_2 = l_5^{\rm NP1}$.
For simplicity, the color factors $C_F=4/3$, $C_A=3$, $T_R=1/2$ and $\mu = m_H$ have been substituted. 
Note that the above result is real, although there are $i\pi$ factors in some terms.
The imaginary part of the $G$ function can be fixed by assigning $w+i0^+$.
We also provide the analytical results in the auxiliary files submitted along this paper.

\section{Asymptotic expansion}
\label{sec:asyexp}

\subsection{Small $m_b$ limit of $\Gamma_{H \rightarrow b\bar{b}}$}
With the full analytical expressions at hand, it is interesting to study the asymptotic behaviors of ${\Delta}^{i,C_1C_1}_{b\bar{b}}$ and ${\Delta}^{C_1C_2}_{i,b\bar{b}}$. Firstly we consider the limit of $z\rightarrow \infty$, corresponding to  $m_b\to 0$. 
${\Delta}^{C_1C_1}_{1,b\bar{b}}$ in this limit is given by
\begin{align}
&{\Delta}^{C_1C_1}_{1,b\bar{b}}|_{z \rightarrow \infty} = \frac{m_H^3}{\pi v^2}C_AC_F
\bigg[\frac{1}{6}  \log (z)-\frac{7}{12}+\frac{3}{z}\bigg]
+\mathcal{O}(z^{-2}).
\end{align}
The logarithm embodies the collinear divergence of a gluon splitting to a bottom quark pair in the massless limit.
The bottom quark mass serves as a regulator.
If one considers the decay to massless bottom quarks, the decay rate is divergent, and thus the decay width of a Higgs boson to massless bottom quarks is not well defined.
In such cases, one has to consider the decay to all partons.

The interference contribution in the small mass limit becomes 
\begin{align}
&{\Delta}^{C_1C_2}_{1,b\bar{b}}|_{z \rightarrow \infty} = \frac{m_Hm_b\overline{m_b}(\mu)}{\pi v^2}C_AC_F\bigg[-\frac{1}{8}\log^2(z)-\frac{3}{4}\log\left(\frac{\mu ^2}{m_H^2}\right)
+\frac{\pi^2}{8}-\frac{19}{8}
\nonumber\\&
+\frac{1}{2}\frac{\log ^2(z)}{z}+2\frac{\log (z)}{z}+\frac{9}{2 z} \log \left(\frac{\mu ^2}{m_H^2}\right)-\frac{\pi ^2}{2 z}+\frac{15}{2 z}\bigg]+\mathcal{O}(z^{-2}),
\label{eq:deltac1c2}
\end{align}
where the double logarithm dominates the contribution over the other terms. 
This kind of double logarithm is induced by soft massive quarks.
It differs from the traditional Sudakov double logarithm, which is induced by soft gluons. 
The all-order structure of this double logarithm is still unclear.
Similar double logarithms appear in the result of the $H\to \gamma\gamma$ decay via a bottom quark loop, and resummation of the double logarithms to all orders has been realized in diagrammatic analyses \cite{Kotsky:1997rq,Liu:2017vkm} or soft-collinear effective theory \cite{Liu:2019oav,Wang:2019mym}. 
However, the methods can not be directly applied here because the corrections of real emissions, which are not needed in $H\to \gamma\gamma$,  have to be included.  
We leave such an interesting topic to future work.
Below we present the higher-order corrections,
\begin{align}
&\tilde{\Delta}^{C_1C_2}_{2,b\bar{b}}|_{z \rightarrow \infty} = \frac{m_Hm_b\overline{m_b}(\mu)}{\pi v^2}\times\nonumber\\
&\bigg(
C_A^2C_F\Big[
-\frac{1}{96}\log ^4(z)+\frac{5}{288}\log ^3(z)+\frac{\pi ^2}{24} \log ^2(z)-\frac{11}{48}\log \left(\frac{\mu ^2}{m_H^2}\right)\log ^2(z)
\nonumber\\&
-\frac{601}{576}\log ^2(z)+\frac{\zeta (3) }{4}\log (z)+\frac{7\pi ^2 }{96}\log (z)+\frac{125}{192}\log (z)-\frac{11}{16}\log ^2\left(\frac{\mu ^2}{m_H^2}\right)
\nonumber\\&
+\frac{11 \pi ^2 }{48}\log \left(\frac{\mu ^2}{m_H^2}\right)
-\frac{51}{8}\log \left(\frac{\mu ^2}{m_H^2}\right)
-\frac{\pi ^4}{40}+\frac{25 \zeta (3)}{12}+\frac{673 \pi ^2}{576}-\frac{19225}{1152}
\nonumber\\&
+\frac{1}{96}\frac{\log ^4(z)}{z}+\frac{2}{9}\frac{\log ^3(z)}{z}+\frac{11}{12}\log \left(\frac{\mu ^2}{m_H^2}\right)\frac{\log ^2(z)}{z}+\frac{\pi ^2}{48}\frac{\log ^2(z)}{z}-\zeta (3) \frac{\log (z)}{z}
\nonumber\\&
-\frac{13 \pi ^2 }{12}\frac{\log (z)}{z}+\frac{103 }{36}\frac{\log ^2(z)}{z}+
\frac{11}{3}\log \left(\frac{\mu ^2}{m_H^2}\right)\frac{\log (z)}{z}+\frac{82}{9}\frac{\log (z)}{z}
\nonumber\\&
+\frac{33}{8 z}\log ^2\left(\frac{\mu ^2}{m_H^2}\right)-\frac{11 \pi ^2}{12 z}\log \left(\frac{\mu ^2}{m_H^2}\right)
+\frac{207}{8 z}\log \left(\frac{\mu ^2}{m_H^2}\right)+\frac{11 \pi ^4}{160 z}
\nonumber\\&
-\frac{55 \zeta (3)}{12 z}
-\frac{185 \pi ^2}{72 z}+\frac{4001}{96 z}
\Big]
\nonumber\\&
+C_AC_F^2\Big[
\frac{1}{64}\log ^4(z)-\frac{5}{32}\log ^3(z)-\frac{3}{32}\log \left(\frac{\mu ^2}{m_H^2}\right)\log ^2(z)-\frac{\pi ^2}{96} \log ^2(z)
+\frac{3}{4}\log ^2(z)
\nonumber\\&
-\frac{3 \zeta (3)}{2} \log (z)-\frac{\pi ^2}{96} \log (z)+\frac{9}{16}\log \left(\frac{\mu ^2}{m_H^2}\right) \log (z) -\frac{23}{32}\log (z)
-\frac{9}{16}\log ^2\left(\frac{\mu ^2}{m_H^2}\right)
\nonumber\\&
+\frac{3\pi ^2 }{32}\log \left(\frac{\mu ^2}{m_H^2}\right)-\frac{141}{32}\log \left(\frac{\mu ^2}{m_H^2}\right)
+\frac{\pi ^4}{64}+\frac{15 \zeta (3)}{4}-\frac{\pi ^2}{6}-\frac{427}{64}
\nonumber\\&
-\frac{1}{16}\frac{\log ^4(z)}{z}-\frac{1}{24}\frac{\log ^3(z)}{z}+\frac{\pi ^2}{24}\frac{\log ^2(z)}{z}
+\frac{3}{8}\log \left(\frac{\mu ^2}{m_H^2}\right)\frac{\log ^2(z)}{z}-\frac{33}{8}\frac{\log ^2(z)}{z}
\nonumber\\&
+6 \zeta (3)\frac{\log (z)}{z}+\frac{41 \pi ^2}{24}\frac{\log (z)}{z}-\frac{69}{8} \log \left(\frac{\mu ^2}{m_H^2}\right)\frac{\log (z)}{z}
-\frac{9}{2}\frac{\log (z)}{z}+\frac{27}{8 z} \log ^2\left(\frac{\mu ^2}{m_H^2}\right)
\nonumber\\&
-\frac{3 \pi ^2}{8 z}\log \left(\frac{\mu ^2}{m_H^2}\right)+\frac{63}{4 z} \log \left(\frac{\mu ^2}{m_H^2}\right)
-\frac{\pi ^4}{16 z}-\frac{25 \zeta (3)}{z}-\frac{43 \pi ^2}{24 z}+\frac{979}{32 z}
\Big]\nonumber\\&
+C_AC_Fn_l\Big[\frac{1}{72}\log ^3(z)+\frac{5}{72}\log ^2(z)-\frac{\pi ^2}{24}\log (z)+\frac{1}{24} \log \left(\frac{\mu ^2}{m_H^2}\right)\log ^2(z)
\nonumber\\&
+\frac{7}{48} \log (z)+\frac{1}{8}\log ^2\left(\frac{\mu ^2}{m_H^2}\right)-\frac{\pi ^2}{24}\log \left(\frac{\mu ^2}{m_H^2}\right)
+\log \left(\frac{\mu ^2}{m_H^2}\right)-\frac{\zeta (3)}{3}-\frac{\pi ^2}{9}+\frac{77}{36}
\nonumber\\&
-\frac{1}{18}\frac{\log ^3(z)}{z}-\frac{1}{6}\log \left(\frac{\mu ^2}{m_H^2}\right)\frac{\log ^2(z)}{z}-\frac{5}{18}\frac{\log ^2(z)}{z}
+\frac{\pi ^2}{6}\frac{\log (z)}{z}
\nonumber\\&
-\frac{2}{3}\log \left(\frac{\mu ^2}{m_H^2}\right)\frac{\log (z)}{z}-\frac{16}{9}\frac{\log (z)}{z}
+\frac{\pi ^2}{6 z}\log \left(\frac{\mu ^2}{m_H^2}\right)
-\frac{3}{4 z}\log ^2\left(\frac{\mu ^2}{m_H^2}\right)
\nonumber\\&
-\frac{15}{4 z}\log \left(\frac{\mu ^2}{m_H^2}\right)+\frac{4 \zeta (3)}{3 z}+\frac{13 \pi ^2}{36 z}-\frac{241}{48 z}
\Big]
\nonumber\\&
+C_AC_F\Big[
\frac{5}{72}\log ^3(z)+\frac{1}{24}\log \left(\frac{\mu ^2}{m_H^2}\right)\log ^2(z)-\frac{19}{144}\log ^2(z)-\frac{7 \pi ^2}{72}\log (z)
\nonumber\\&
+\frac{3}{16}\log (z)+\frac{1}{8} \log ^2\left(\frac{\mu ^2}{m_H^2}\right)-\frac{\pi ^2}{24}\log \left(\frac{\mu ^2}{m_H^2}\right)+\log \left(\frac{\mu ^2}{m_H^2}\right)+\frac{23 \pi ^2}{432}+\frac{43}{16}
\nonumber\\&
-\frac{5}{18}\frac{\log ^3(z)}{z}-\frac{1}{6}\log \left(\frac{\mu ^2}{m_H^2}\right)\frac{\log ^2(z)}{z}+\frac{4}{9}\frac{\log ^2(z)}{z}
+\frac{7 \pi ^2}{18}\frac{\log (z)}{z}
\nonumber\\&
-\frac{2}{3}\log \left(\frac{\mu ^2}{m_H^2}\right)\frac{\log (z)}{z}
-\frac{101}{36}\frac{\log (z)}{z}
-\frac{3}{4 z}\log ^2\left(\frac{\mu ^2}{m_H^2}\right)+\frac{\pi ^2}{6 z}\log \left(\frac{\mu ^2}{m_H^2}\right)
\nonumber\\&
-\frac{15}{4 z}\log \left(\frac{\mu ^2}{m_H^2}\right)-\frac{65 \pi ^2}{108 z}-\frac{1811}{144 z}
\Big]
\bigg)+\mathcal{O}(z^{-2})
\label{Ser_bb}
\end{align}
and
\begin{align}
&{\Delta}^{C_1C_2}_{2,b\bar{b}b\bar{b}}|_{z \rightarrow \infty} = \frac{m_Hm_b\overline{m_b}(\mu)}{\pi v^2}\times\nonumber\\
&\bigg(
\left(C_A^2C_F-2C_AC_F^2\right)\Big[\frac{1}{192}\log ^4(z)-\frac{1}{16}\log ^3(z)-\frac{\pi ^2}{96}\log ^2(z)
+\frac{7}{16}\log ^2(z)
\nonumber\\&
+\frac{\pi ^2}{16}\log (z)
-\frac{5}{4}\log (z)
-\frac{\zeta (3)}{4}\log (z)
+\frac{19 \pi ^4}{960}+\frac{7 \zeta (3)}{8}-\frac{\pi ^2}{3}+\frac{23}{16}
\nonumber\\&
-\frac{1}{48}\frac{\log ^4(z)}{z}+\frac{1}{6}\frac{\log ^3(z)}{z}+\frac{\pi ^2}{24}\frac{\log ^2(z)}{z}-\frac{13}{8}\frac{\log ^2(z)}{z}
+\zeta (3)\frac{\log (z)}{z}-\frac{\pi ^2}{12}\frac{\log (z)}{z}
\nonumber\\&
+\frac{9}{2}\frac{\log (z)}{z}
-\frac{19 \pi ^4}{240 z}
-\frac{9 \zeta (3)}{2 z}+\frac{7 \pi ^2}{8 z}-\frac{7}{2 z}\Big]
\nonumber\\&
+C_AC_F\Big[
-\frac{1}{36}\log ^3(z)+\frac{19}{144}\log ^2(z)+\frac{\pi ^2}{18}\log (z)-\frac{17}{48}\log (z)-\frac{\zeta (3)}{2}-\frac{41 \pi ^2}{432}+\frac{25}{96}
\nonumber\\&
+\frac{1}{9}\frac{\log ^3(z)}{z}-\frac{11}{18}\frac{\log ^2(z)}{z}-\frac{2 \pi ^2}{9}\frac{\log (z)}{z}+\frac{31}{18}\frac{\log (z)}{z}+\frac{2 \zeta (3)}{z}+\frac{37 \pi ^2}{54 z}+\frac{25}{18 z}
\Big]
\bigg)+\mathcal{O}(z^{-2}),
\label{Ser_bbbb}
\end{align}
for the two and four bottom quark final states, respectively.
Both contributions exhibit double logarithmic enhancement, i.e., the $\als^n \log^{2n}(z)$ expansion pattern.
These terms break the perturbative convergence of the higher-order corrections and have to be resumed to all orders to provide reliable theoretical predictions.
The above analytic results can be used to check the consistency of the resummation formula in the future.

Combining the two corrections in eq. (\ref{Ser_bb}) and eq. (\ref{Ser_bbbb}), 
we get the leading logarithm of ${\Delta}^{C_1C_2}_{2,b\bar{b}}$,
\begin{align}
-\frac{m_Hm_b\overline{m_b}(\mu)}{8\pi v^2}C_A C_F\log^2(z) \times \frac{1}{24}\left(C_A-C_F\right)\log^2(z).
\end{align}
This color structure distinguishes it from the Sudakov double logarithms 
and shares the same features as the results for the quark-gluon splitting function \cite{Vogt:2010cv}, $Hb\bar{b}$/$Hgg$ form factors \cite{Liu:2017vkm,Liu:2022ajh}, and off-diagonal “gluon” thrust \cite{Moult:2019uhz,Beneke:2020ibj,Beneke:2022obx}.

To provide the full $\mathcal{O}(\als^3)$ corrections, 
we also need the analytical expression of the third order correction in $\Gamma^{C_2 C_2}_{H\rightarrow b\bar{b}}$ \cite{Gorishnii:1990zu,Chetyrkin:1996sr,Baikov:2005rw,Herzog:2017dtz}, 
\begin{align}
&\Delta^{C_2C_2}_{3,b\bar{b}}|_{z\to \infty} = \frac{m_H\overline{m_b}(\mu)^2}{512v^2\pi }C_A\times 
\nonumber\\&
\bigg(C_AC_F^2\left[
\frac{100 \zeta (5)}{3}-\frac{4658 \zeta (3)}{3}-\frac{3430 \pi ^2}{27}+\frac{3894493}{972}
\right]
\nonumber\\&
+C_A^2C_F\left[
580 \zeta (5)-2178 \zeta (3)-\frac{766 \pi ^2}{3}+\frac{13153}{3}
\right]
\nonumber\\&
+C_F^3\left[
360 \zeta (5)-956 \zeta (3)-108 \pi ^2+\frac{23443}{12}
\right]
\nonumber\\&
+C_AC_Fn_f\left[
-\frac{80 \zeta (5)}{3}+\frac{4 \pi ^4}{15}-\frac{704 \zeta (3)}{3}-\frac{1142 \pi ^2}{27}+\frac{267800}{243}
\right]
\nonumber\\&
+C_F^2n_f\left[
160 \zeta (5)-\frac{4 \pi ^4}{15}-520 \zeta (3)-\frac{130 \pi ^2}{3}+\frac{2816}{3}\right]
\nonumber\\&
+C_Fn_f^2\left[
-16 \zeta (3)-\frac{88 \pi ^2}{27}+\frac{15511}{243}
\right]\bigg) +\mathcal{O}(z^{-1}),
\label{DeltaC2C2_3bb}
\end{align}
where $\mu = m_H$ have been substituted.
We see that there is no logarithmic enhancement in this part.
The $\mathcal{O}(z^{-1})$ term has been computed at $\als^2$ in \cite{Chetyrkin:1995pd, Harlander:1997xa, Wang:2023xud}, 
and is found to be less than $1\%$ of the $\mathcal{O}(z^{0})$ contribution.
The result of $\mathcal{O}(z^{-1})$ term at $\als^3$ is expected to be also a few percent of the $\mathcal{O}(z^{0})$ term at this order, and thus completely negligible.

Combining all the above expressions with the Wilson coefficients in eq. (\ref{C1C2}), and converting the on-shell $m_b$ to $\overline{m_b}$ in the  $\overline{\text{MS}}$ scheme via the relation 
\begin{align}
\overline{m_b}(\mu) = m_b\left( 1 - \left(\frac{\als}{\pi}\right)C_F\Big[1+\frac{3}{4}\log \left(\frac{\mu ^2}{m_b^2}\right)\Big]+\mathcal{O}(\als^2)\right)\,,   
\end{align}
we obtain a compact result of $\Gamma_{H\rightarrow b\bar{b}}$ up to $\mathcal{O}(\als^{3})$, 
\begin{align}
\Gamma_{H\rightarrow b\bar{b}} &= 
\frac{3m_H\overline{m_b}^2}{8v^2\pi }\bigg\{1+\left(\frac{\als}{\pi}\right)\frac{17}{3}
\nonumber\\&\quad
+\left(\frac{\als}{\pi}\right)^2\Big[\frac{1}{9}\log^2(\overline{z})-\frac{2}{3}\log (x)-\frac{97 \zeta (3)}{6}-\frac{17\pi ^2}{12}+\frac{9235}{144}\Big]
\nonumber\\&\quad
+\left(\frac{\als}{\pi}\right)^3\Big[\frac{5}{648}\log ^4(\overline{z})
+\frac{59}{324}\log ^3(\overline{z})-\frac{31\pi ^2}{324}\log ^2(\overline{z})
+\frac{989}{648}\log ^2(\overline{z})
\nonumber\\&\quad
+\frac{32 \zeta (3) }{27}\log (\overline{z})
-\frac{41\pi ^2}{324}\log (\overline{z})+\frac{137}{216}\log (\overline{z})-\frac{23}{18}\log^2 (x)
-\frac{49}{6}\log (x)
\nonumber\\&\quad
+\frac{1945 \zeta (5)}{36}
-\frac{13 \pi ^4}{3240}
-\frac{81239 \zeta (3)}{216}
-\frac{81239 \zeta (3)}{216}-\frac{22291 \pi ^2}{648}+\frac{37434709}{46656}\Big]\bigg\}
\nonumber\\&\quad
+\frac{m_H^3}{v^2\pi }\left(\frac{\als}{\pi}\right)^3\Big[\frac{\log(\overline{z})}{216}-\frac{7}{432}\Big]
+\mathcal{O}(\overline{z}^{-1})+\mathcal{O}(x)+\mathcal{O}(\als^4)
\label{eq:bbexp}
\end{align}
with $\overline{z} = m_H^2/\overline{m_b}^2$ and $x = m_H^2/m_t^2$. 
The color factors $C_F=4/3$, $C_A=3$, $T_R=1/2$, $n_f=5$, $n_l=4$ and $\mu = m_H$ have been applied after we checked the scale independence  of the expression up to $\mathcal{O}(\als^3)$.
The above expression up to $\mathcal{O}(\als^2)$ agrees with the results in \cite{Chetyrkin:1995pd,Larin:1995sq}. 
The $\mathcal{O}(\als^3)$ correction is new.
The large logarithmic terms, $\log^i(\overline{z}), i=1,\cdots,4$, in the curly bracket provide significant corrections to the decay width.
The last term in the above equation is of order $\als^3$, but is power enhanced by $\bar{z}\log(\overline{z})$ compared to the LO decay width.
If this term dominates the contribution, it would be hard to extract the value of bottom quark mass from the decay width because of its logarithmic dependence on $m_b$.
Fortunately, due to the small coefficient, this term does not play a significant role.

We also calculate the partial decay width of $\Gamma_{H\rightarrow gg}$ in the small $m_b$ limit explicitly in a similar way, and the result is given by
\begin{align}
\Gamma_{H\rightarrow gg} &= 
\frac{m_H\overline{m_b}^2}{v^2\pi }\bigg\{
\left(\frac{\als}{\pi}\right)^2\Big[-\frac{1}{24}\log^2(\overline{z})+\frac{\pi^2}{24}+\frac{1}{6}\Big]
\nonumber\\&\quad
+\left(\frac{\als}{\pi}\right)^3\Big[-\frac{5}{1728}\log ^4(\overline{z})
-\frac{59}{864}\log ^3(\overline{z})+\frac{31\pi ^2}{864}\log ^2(\overline{z})
-\frac{989}{1728}\log ^2(\overline{z})
\nonumber\\&\quad
-\frac{4 \zeta (3) }{9}\log (\overline{z})+\frac{41\pi ^2}{864}\log (\overline{z})-\frac{137}{576}\log (\overline{z})
-\frac{137 \pi ^4}{8640}
-\frac{29 \zeta (3)}{36}
\nonumber\\&\quad
+\frac{1277 \pi ^2}{1728}+\frac{17275}{3456}\Big]\bigg\}
+\frac{m_H^3}{v^2\pi }\bigg\{\left(\frac{\als}{\pi}\right)^2\frac{1}{72}+\left(\frac{\als}{\pi}\right)^3\Big[-\frac{1}{216}\log (\overline{z})+\frac{229}{864}\Big]\bigg\}
\nonumber\\&\quad
+\mathcal{O}(\overline{z}^{-1})+\mathcal{O}(x)+\mathcal{O}(\als^4).
\end{align}
Similar large logarithmic/power enhancements are observed.

Summing up the above two contributions, we obtain the decay width of the Higgs boson to all hadrons,
\begin{align}
\Gamma_{H\rightarrow \text{hadrons}} &= \frac{3m_H\overline{m_b}^2}{8v^2\pi }\bigg\{1+\left(\frac{\als}{\pi}\right)\frac{17}{3}
+\left(\frac{\als}{\pi}\right)^2\Big[-\frac{2}{3}\log(x)-\frac{97 \zeta (3)}{6}-\frac{47 \pi ^2}{36}+\frac{9299}{144}\Big]
\nonumber\\&\quad
+\left(\frac{\als}{\pi}\right)^3\Big[-\frac{23}{18}\log ^2(x)-\frac{49}{6}\log (x)+\frac{1945 \zeta (5)}{36}-\frac{5 \pi ^4}{108}
-\frac{81703 \zeta (3)}{216}
\nonumber\\&\quad
-\frac{10507 \pi ^2}{324}+\frac{38056609}{46656}\Big]
\bigg\}
+\frac{m_H^3}{v^2\pi }\bigg\{\left(\frac{\als}{\pi}\right)^2\frac{1}{72}+\left(\frac{\als}{\pi}\right)^3\frac{215}{864}\bigg\}
\nonumber\\&\quad
+\mathcal{O}(\overline{z}^{-1})+\mathcal{O}(x)+\mathcal{O}(\als^4).
\end{align}
The above result coincides with that in refs. \cite{Chetyrkin:1997vj,Davies:2017xsp} obtained in a different method. 
This is a strong check of our calculation.

\subsection{Threshold limit}
It is also interesting to investigate the threshold limit of bottom quark pair production. 
Expanding the full analytic results in the limit of $ m_H\rightarrow 2m_b$ or $\beta\rightarrow 0$, we have
\begin{align}
&\Delta^{C_1C_1}_{1,b\bar{b}}|_{\beta \rightarrow 0}  = \frac{m_H^3}{\pi v^2}C_AC_F\frac{4}{315}\beta^9+\mathcal{O}(\beta^{10})
\end{align}
and
\begin{align}
&\Delta^{C_1C_2}_{1,b\bar{b}}|_{\beta \rightarrow 0}  = \frac{m_Hm_b\overline{m_b}(\mu)}{\pi v^2}C_AC_F\Big[
-\frac{3}{4} \log \left(\frac{\mu ^2}{m_H^2}\right)-\frac{\log(2)}{2}-1
\Big]\beta^3+\mathcal{O}(\beta^4).
\label{eq:deltac1c2beta0}
\end{align}
It can be seen that the contribution from the $\mathcal{O}_1$ operator is highly suppressed near the threshold. 
For comparison, we also present the result of $\Delta^{C_2C_2}_{1,b\bar{b}}$,
\begin{align}
&\Delta^{C_2C_2}_{1,b\bar{b}}|_{\beta \rightarrow 0}  = \frac{m_H\overline{m_b}(\mu)^2}{\pi v^2}C_AC_F\frac{\pi^2}{16}\beta^2+\mathcal{O}(\beta^{3}) ,
\end{align}
which is $1/\beta$ power enhanced with respect to the LO result due to Coulomb interaction between the bottom quark pair.
The NLO result of $\Delta^{C_1C_2}_{b\bar{b}}$ in the threshold limit is given by
\begin{align}
&\Delta^{C_1C_2}_{2,b\bar{b}}|_{\beta \rightarrow 0}  = \frac{m_Hm_b\overline{m_b}(\mu)}{\pi v^2}
\bigg(
C_AC_F^2\beta^2\pi^2\Big[-\frac{3}{8} \log \left(\frac{\mu ^2}{m_H^2}\right)-\frac{\log(2)}{4}-\frac{1}{2}\Big]
\nonumber\\&
+C_A^2C_F\beta^3\Big[
-\frac{11}{16}\log ^2\left(\frac{\mu ^2}{m_H^2}\right)-\frac{11}{12} \log (2) \log \left(\frac{\mu ^2}{m_H^2}\right)-\frac{185}{48} \log \left(\frac{\mu ^2}{m_H^2}\right)
\nonumber\\&
+\frac{7 \zeta (3)}{16}+\frac{5\pi ^2 \log (2)}{12}+\frac{5 \pi ^2}{36}-\frac{11 \log ^2(2)}{12}-\frac{61 \log (2)}{72}-\frac{3557}{576}
\Big]
\nonumber\\&
+C_AC_F^2\beta^3\Big[
-\frac{9}{16}\log ^2\left(\frac{\mu ^2}{m_H^2}\right)-\frac{3\log (2)}{2}\log \left(\frac{\mu ^2}{m_H^2}\right)-\frac{15}{16}\log \left(\frac{\mu ^2}{m_H^2}\right)
\nonumber\\&
+\frac{13 \zeta (3)}{16}-\frac{7 \pi ^2 \log (2)}{12}+\frac{209 \pi ^2}{576}-\frac{3 \log ^2(2)}{4}-\frac{61 \log (2)}{24}-\frac{37}{64}
\Big]
\nonumber\\&
+C_AC_Fn_l\beta^3\Big[
\frac{1}{8}\log ^2\left(\frac{\mu ^2}{m_H^2}\right)+\frac{\log (2)}{6}\log \left(\frac{\mu ^2}{m_H^2}\right)+\frac{13}{24}\log \left(\frac{\mu ^2}{m_H^2}\right)
\nonumber\\&
-\frac{\pi ^2}{18}+\frac{\log ^2(2)}{6}+\frac{5 \log (2)}{36}+\frac{229}{288}
\Big]
\nonumber\\&
+C_AC_F\beta^3\Big[
\frac{1}{8}\log ^2\left(\frac{\mu ^2}{m_H^2}\right)+\frac{\log (2)}{6}\log \left(\frac{\mu ^2}{m_H^2}\right)+\frac{13}{24}\log \left(\frac{\mu ^2}{m_H^2}\right)
\nonumber\\&
-\frac{37 \pi ^2}{288}-\frac{\log ^2(2)}{6}+\frac{13 \log (2)}{12}+\frac{281}{288}
\Big]
\bigg)+\mathcal{O}(\beta^4).
\end{align}
Compared to the LO result in (\ref{eq:deltac1c2beta0}), it exhibits the same power enhancement as predicted by the Coulomb Green function; see eq. (5.15) of ref. \cite{Wang:2023xud}.

\begin{figure}
    \centering
    \includegraphics[width=0.65\linewidth]{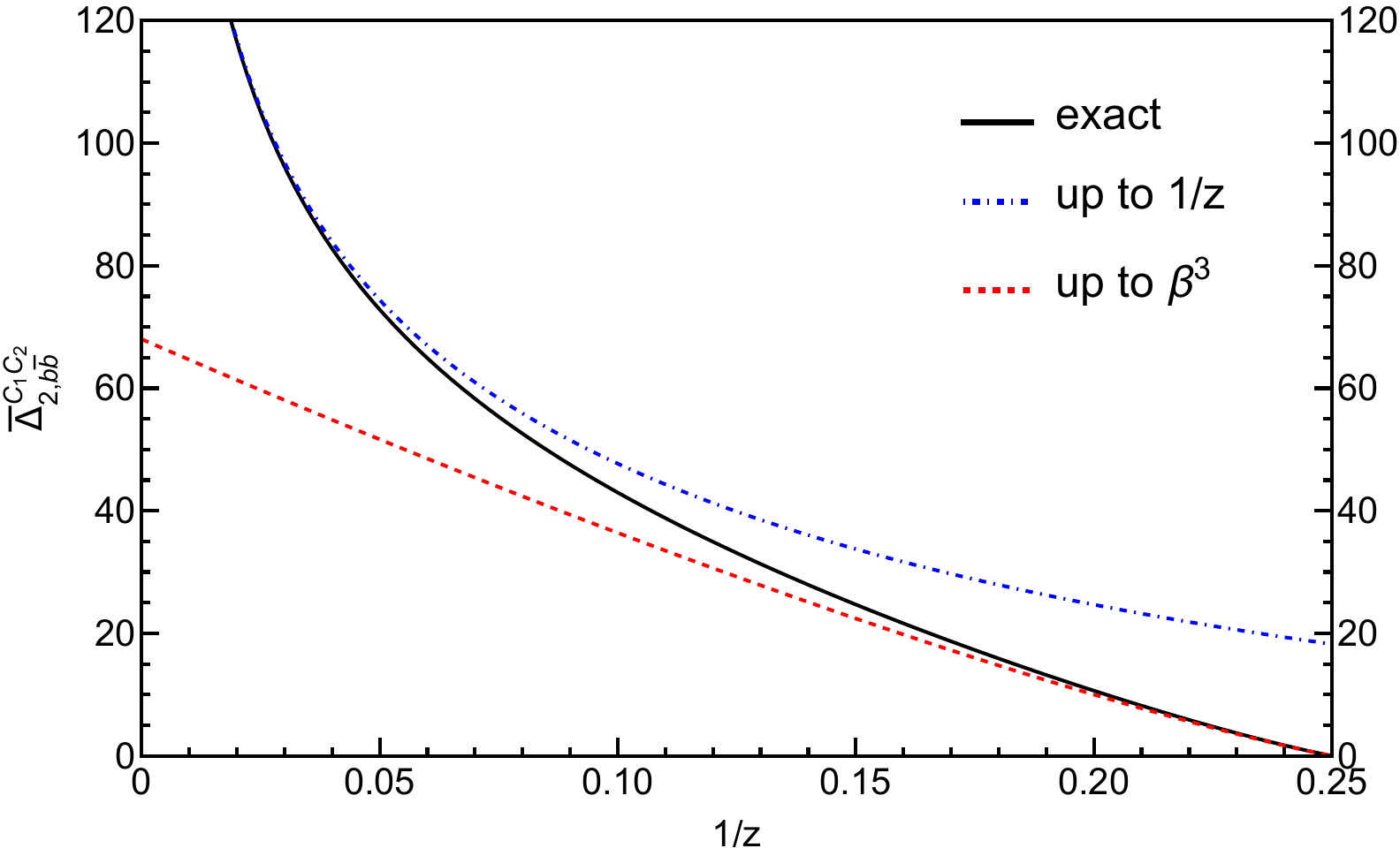}
    \caption{Comparison between the exact result of $\overline{\Delta}^{C_1C_2}_{2,b\bar{b}}$ (black line) and its expansion around $z \rightarrow \infty$ (dot-dashed blue line) or $z \rightarrow 1/4$
    (dashed red line).}
    \label{Difference}
\end{figure}

To illustrate the expansions around the small mass and threshold limits, we show in figure  \ref{Difference} the numerical results of $\overline{\Delta}^{C_1C_2}_{2,b\bar{b}}$, which is defined by
\begin{align}
\overline{\Delta}^{C_1C_2}_{2,b\bar{b}} \equiv {\Delta}^{C_1C_2}_{2,b\bar{b}}/\left(\frac{m_Hm_b\overline{m_b}(\mu)}{\pi v^2} \right).
\end{align}
Here we have divided ${\Delta}^{C_1C_2}_{2,b\bar{b}}$ by its overall coupling to expose the dependence of the rest part on $z$ more clearly.
The small mass expansion  coincides with the exact result for $1/z<0.05$,
while the threshold expansion overlaps with the exact result when $1/z>0.2$.

\section{Numerical result}
\label{sec:num}

To give the numerical result of the decay width of $H\to b\bar{b}$,
we take the input parameters
\begin{align}
    \overline{m_b}\,(\overline{m_b}) &= 4.18 ~{\rm GeV},  & m_b & = 5.07~{\rm GeV}, & m_H & = 125.09 ~{\rm GeV},  \\
     m_t & = 172.69 ~{\rm GeV}, 
    & \als(m_Z) & = 0.1181, & G_F & = 1.166378 \times 10^{-5}~{\rm GeV}^{-2}, \nn
    \label{eq:input}    
\end{align}
where the on-shell value of $m_b$ is obtained from $\overline{m_b}(\overline{m_b})$ in the $\overline{\rm MS}$ scheme via the four-loop mass relation \cite{Broadhurst:1991fy,Gray:1990yh,Marquard:2015qpa,Marquard:2016dcn}. The package {\tt RunDec} \cite{Chetyrkin:2000yt,Herren:2017osy} was used to obtain $\overline{m_b}$ at other scales, e.g., 
$\overline{m_b}\,(m_H/2) = 2.95631$ GeV, $\overline{m_b}\,(m_H) = 2.78425$ GeV and $\overline{m_b}\,(2m_H) = 2.63908$ GeV. 
The strong coupling at several typical scales is evaluated to be $\als\,(m_H/2) = 0.125257$, $\als\,(m_H) = 0.112715$ and $\als\,(2m_H) = 0.102501$.

In table \ref{bbX1X2}, we show the QCD corrections induced by different combinations of operators.
The LO and NLO results are obtained only by considering the Yukawa coupling, exhibiting a NLO correction of 20\%.
The NNLO correction increases the NLO decay width further by 4\%, and begins to receive contributions from the Higgs-gluon-gluon vertices, which account for one third of the $\mathcal{O}(\als^2)$ correction.
The N$^3$LO correction gives rise to 1\% improvement with respect to the NNLO result,
much larger than the estimated contribution, about $0.2\%$, from missing higher-order corrections based on the calculation with massless bottom quarks \cite{Spira:2019iec}.
It is interesting to observe the large higher-order correction in the interference contribution $\Gamma^{C_1C_2}_{Hb\bar{b}}$, about $55\%$.
This is due to the large logarithms $\ln(m_H^2/m_b^2)$ as shown in the analytic result in eq. (\ref{eq:bbexp}).

\begin{table}[H]
	\centering
	\scalebox{1.0}{
		\begin{tabular}{clccc}
			\toprule
            &
			[MeV]&
			$\mu = \frac{1}{2}m_H$\quad &
			$\mu = m_H$\quad &
			$\mu = 2m_H$ \quad\\
			\midrule
            $\mathcal{O}(\als^0)$&
			$\Gamma^{C_2C_2}_{Hb\bar{b}}$&
			2.1314&
			1.8905&
            1.6985
			\\
			\midrule
            $\mathcal{O}(\als^1)$&
			$\Gamma^{C_2C_2}_{Hb\bar{b}}$&
			0.25813&
			0.39409&
			0.47563\\
            \midrule
            \multirow{2.5}{*}{$\mathcal{O}(\als^2)$}&
			$\Gamma^{C_2C_2}_{Hb\bar{b}}$&
			-0.0043084&
			0.076819&
			0.143670\\
			\cmidrule{2-5}&
			$\Gamma^{C_1C_2}_{Hb\bar{b}}$&
            0.027078&
			0.024746&
			0.022608\\
			\midrule
            \multirow{4}{*}{$\mathcal{O}(\als^3)$}&
            $\Gamma^{C_2C_2}_{Hb\bar{b}}$&
            -0.015360&
			0.0048336&
			0.038198\\
   		\cmidrule{2-5}&
   		$\Gamma^{C_1C_2}_{Hb\bar{b}}$&
            0.010314&
            0.013585&
			0.015170\\
      	\cmidrule{2-5}
            &$\Gamma^{C_1C_1}_{Hb\bar{b}}$&
            0.0088676&
			0.0064617&
			0.0048595\\
			\bottomrule
		\end{tabular}}
	\caption{ $\Gamma^{C_2C_2}_{Hb\bar{b}}$, $\Gamma^{C_1C_2}_{Hb\bar{b}}$ and $\Gamma^{C_1C_1}_{Hb\bar{b}}$ at different orders of $\als$ in the mixed mass scheme.} 
	\label{bbX1X2}
\end{table}

The above results are obtained in the mixed mass scheme, i.e., the Yukawa coupling and the masses in propagators are renormalized in the $\overline{\rm MS}$ and on-shell schemes, respectively.
This scheme is convenient in organizing the calculation.
For phenomenological study, it is more relevant to use a unified scheme.
Given our analytical results, it is straight forward 
to derive the result in either the $\overline{\rm MS}$ or the on-shell scheme.
We show the corresponding numerical results of different higher-order corrections in table \ref{bbX1X2twoschemes}.
The results in the $\overline{\rm MS}$ scheme are very similar to those in the mixed scheme.
The two schemes differ in the renormalization scheme of the mass in propagators.
This would affect only the subleading power 
 contribution in the result of $\Gamma^{C_2C_2}_{Hb\bar{b}}$ up to $\mathcal{O}(\als^2)$, which is highly suppressed by $m_b^2/m_H^2$.
At  $\mathcal{O}(\als^3)$, the difference arises at leading power, but is still small.
The change in the result of $\Gamma^{C_1C_2}_{Hb\bar{b}}$ is obvious because it is proportional to the propagator mass; see eq. (\ref{eq:deltac1c2}).
However, the induced change in the total decay width is not significant due to the suppression of $\als^2$.
The scale uncertainty in the $\overline{\rm MS}$ scheme is $3\%$ and $1\%$ at NNLO and N$^3$LO, respectively.
We provide the result of the decay width at N$^3$LO,
\begin{align}
    \Gamma_{H\rightarrow b\bar{b}}^{\rm N^3LO ~QCD}\left(\overline{\rm MS}\right) = 2.410 ^{+0.007}_{-0.017} {~\rm MeV}.
\end{align}
This accuracy reaches the same level of the experimental expectation.
The corrections from finite top quark mass effects are only about $0.001$ MeV \cite{Chetyrkin:1995pd, Larin:1995sq} and thus can be neglected.
Including the NLO EW corrections \cite{Mihaila:2015lwa}, we obtain
\begin{align}
    \Gamma_{H\rightarrow b\bar{b}}^{\rm N^3LO ~QCD+NLO ~EW}\left(\overline{\rm MS}\right) = 2.382 ^{+0.007}_{-0.017} {~\rm MeV}.
\end{align}

The results in the on-shell scheme exhibit large higher-order corrections.
The LO result is 3.3 (which is just about $m_b^2/\overline{m_b}^2(m_H)$) times of the one in the $\overline{\rm MS}$ scheme.
The decay width is decreased by $-35\%$ after including the NLO correction.
The NNLO correction reduces the decay width further by $-22\%$.
There is still a correction of $-10\%$ at N$^3$LO.
This poor convergence is caused by the fact that the on-shell mass, as a long-distance definition, suffers from infra-red renormalons \cite{Beneke:1998ui}.

\begin{figure}[ht]
	\centering
	\includegraphics[width=0.7\linewidth]{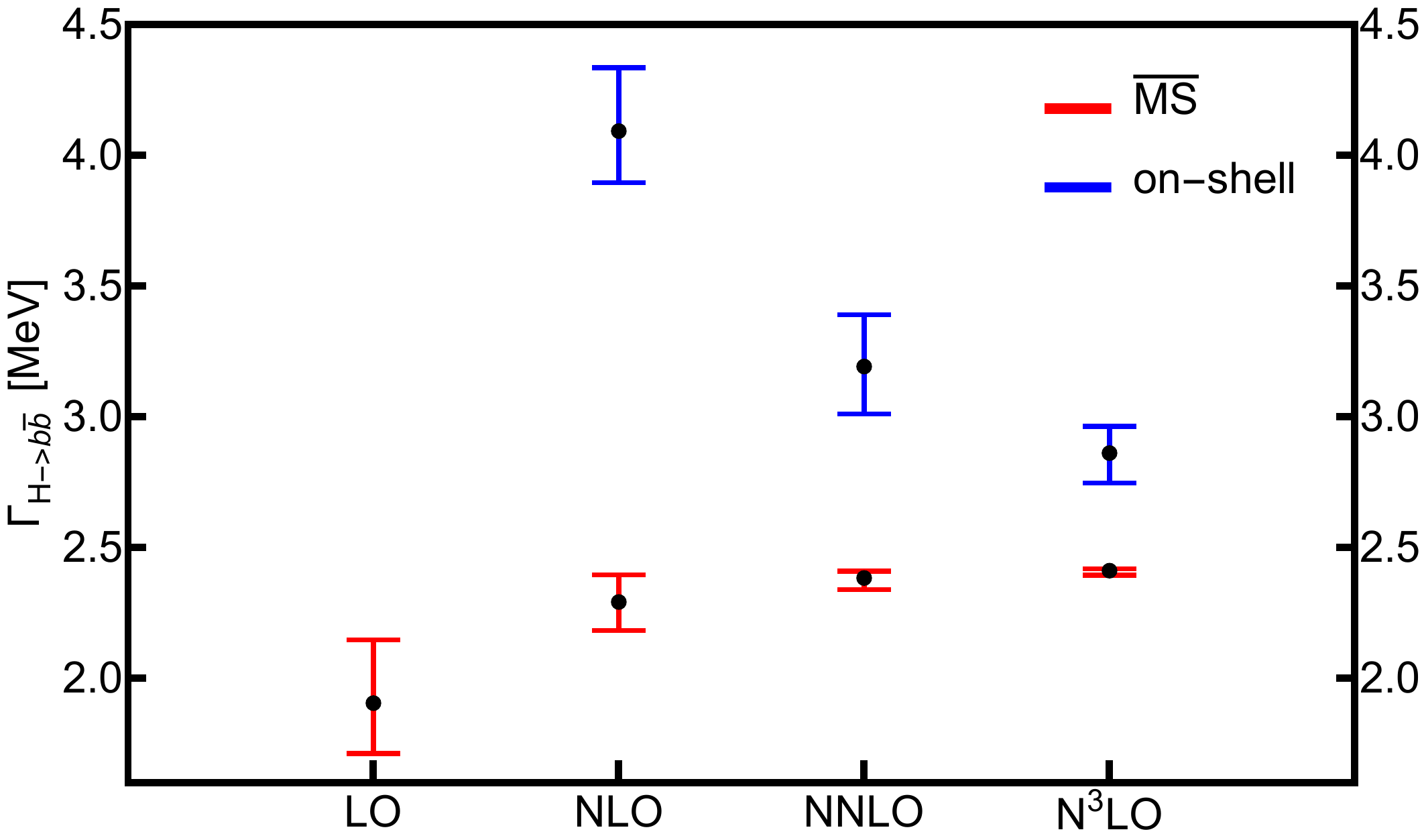}
\caption{Comparison of the results in the $\overline{\rm MS}$ and the on-shell schemes. The error bars denote the scale uncertainties.}
\label{scheme}
\end{figure}

The comparison between the results in the two schemes is shown in figure  \ref{scheme}.
As more high-order corrections are included, the scale uncertainties become notably smaller, and the difference between the two renormalization schemes is also reduced.

\begin{table}[H]
	\centering
	\scalebox{1.0}{
		\begin{tabular}{clccc}
			\toprule
            &
			[MeV]&
			$\mu = \frac{1}{2}m_H$\quad &
			$\mu = m_H$\quad &
			$\mu = 2m_H$ \quad\\
			\midrule
            \multirow{5}{*}{$\Gamma_{H\rightarrow b\bar{b}}\left(\overline{\rm MS}\right)$}&
			$\mathcal{O}(\als^0)$&
            2.1454&
			1.9036&
			1.7108\\
			\cmidrule{2-5}
   		&$\mathcal{O}(\als^1)$&
            0.24806&
			0.38682&
			0.47051\\
			\cmidrule{2-5}
      	&$\mathcal{O}(\als^2)$&
            0.014742&
			0.091773&
			0.15580\\
			\cmidrule{2-5}
            &$\mathcal{O}(\als^3)$&
            0.0092203&
			0.028117&
			0.055816\\
			\midrule   
            \multirow{5}{*}{$\Gamma_{H\rightarrow b\bar{b}}\left({\rm OS}\right)$}&
   		$\mathcal{O}(\als^0)$&
            6.2687&
            6.2687&
			6.2687\\
      	\cmidrule{2-5}
            &$\mathcal{O}(\als^1)$&
            -2.4192&
			-2.1770&
			-1.9797\\
            \cmidrule{2-5}
      	&$\mathcal{O}(\als^2)$&
            -0.85590&
            -0.90061&
			-0.91641\\
      	\cmidrule{2-5}
            &$\mathcal{O}(\als^3)$&
            -0.23550&
			-0.33107&
			-0.39831\\
			\bottomrule
		\end{tabular}}
	\caption{The result of the decay width in the $\overline{\rm MS}$ or the on-shell schemes.} 
	\label{bbX1X2twoschemes}
\end{table}

\section{Conclusion}
\label{sec:conclusion}
We have calculated the analytic result of the dominant decay channel of the Higgs boson, $H\to b\bar{b}$,  at $\mathcal{O}(\alpha_s^3)$.
For the process induced by the bottom quark  Yukawa coupling $y_b$, we retain full bottom quark mass dependence up to NNLO but use the result with massless final-state bottom quarks for the $\mathcal{O}(\alpha_s^3)$ correction. 
The higher-power mass effects in this process are tiny due to the small Taylor expansion parameter $m_b^2/m_H^2$.
For the process induced by the top quark Yukawa coupling $y_t$, we perform the calculation always with finite $m_b$.
We consider the contributions from both two and four massive bottom quarks in the final states.
The result of the former can be expressed in terms of   multiple polylogarithms, while the latter needs elliptic integrals.
The asymptotic expansion of the decay width for $H\to b\bar{b}$ in the small $m_b$ limit shows double logarithmic enhancement from $\mathcal{O}(\alpha_s^2)$.
The coefficient of the double logarithm at $\mathcal{O}(\alpha_s^3)$ is proportional to $C_A-C_F$, which is a typical color structure in the subleading power resummation with soft quarks.
Our analytic results can be used to check the resummation formula in future.
The $\mathcal{O}(\alpha_s^3)$ correction increases the NNLO decay rate by $1\%$, much larger than expectation from naive power counting of $\alpha_s$.
This improvement is mainly due to the large logarithms as mentioned above.
Combining our result with the NLO EW corrections, we provide the most accurate prediction for the $H\to b\bar{b}$ decay with a scale uncertainty of $1\%$ in the $\overline{\rm MS}$ scheme of the bottom quark mass renormalization.
The results in the on-shell scheme exhibit poor convergence.
However, the difference between the two schemes reduces as more higher-order corrections are included.

\section*{Acknowledgements}
We thank Lorenzo Tancredi for discussions. This work was supported in part by the National Science Foundation of China (grant Nos. 12405117, 12005117, 12321005, 12375076), the Taishan Scholar Foundation of Shandong province (tsqn201909011) and the Deutsche Forschungsgemeinschaft (DFG, German Research Foundation) under Germany's Excellence Strategy – EXC-2094 – 390783311. J. W. and X. W. thank the Munich Institute for Astro-, Particle and BioPhysics (MIAPbP), which is funded by the same DFG grant, for hospitality during part of the work.

\appendix
\section{Topologies of the master integrals}
\label{sec:topo}

\begin{figure}[H]
	\centering
	\begin{minipage}{0.2\linewidth}
		\centering
		\includegraphics[width=1\linewidth]{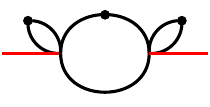}
		\caption*{$\M^{\text{NP1}}_{1}$}
	\end{minipage}
	\begin{minipage}{0.2\linewidth}
		\centering
		\includegraphics[width=1\linewidth]{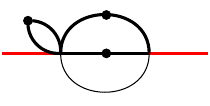}
		\caption*{$\M^{\text{NP1}}_{2}$}
	\end{minipage}
	\begin{minipage}{0.2\linewidth}
		\centering
		\includegraphics[width=1\linewidth]{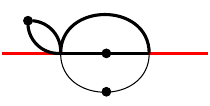}
		\caption*{$\M^{\text{NP1}}_{3}$}
	\end{minipage}
	\centering
	\begin{minipage}{0.2\linewidth}
		\centering
		\includegraphics[width=1\linewidth]{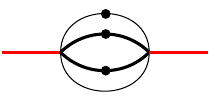}
		\caption*{$\M^{\text{NP1}}_{4}$}
	\end{minipage}
	\begin{minipage}{0.2\linewidth}
		\centering
		\includegraphics[width=1\linewidth]{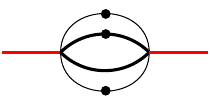}
		\caption*{$\M^{\text{NP1}}_{5}$}
	\end{minipage}
	\begin{minipage}{0.2\linewidth}
		\centering
		\includegraphics[width=1\linewidth]{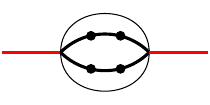}
		\caption*{$\M^{\text{NP1}}_{6}$}
	\end{minipage}
	\begin{minipage}{0.2\linewidth}
		\centering
		\includegraphics[width=1\linewidth]{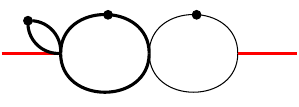}
		\caption*{$\M^{\text{NP1}}_{7}$}
	\end{minipage}
	\begin{minipage}{0.2\linewidth}
		\centering
		\includegraphics[width=1\linewidth]{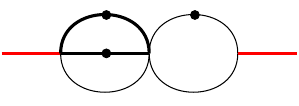}
		\caption*{$\M^{\text{NP1}}_{8}$}
	\end{minipage}
	\begin{minipage}{0.2\linewidth}
		\centering
		\includegraphics[width=1\linewidth]{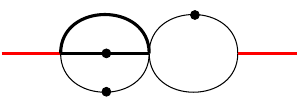}
		\caption*{$\M^{\text{NP1}}_{9}$}
	\end{minipage}
	\begin{minipage}{0.2\linewidth}
		\centering
		\includegraphics[width=1\linewidth]{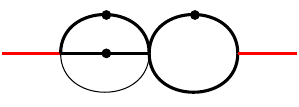}
		\caption*{$\M^{\text{NP1}}_{10}$}
	\end{minipage}
	\begin{minipage}{0.2\linewidth}
		\centering
		\includegraphics[width=1\linewidth]{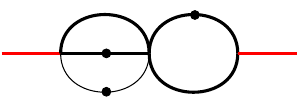}
		\caption*{$\M^{\text{NP1}}_{11}$}
	\end{minipage}
	\begin{minipage}{0.2\linewidth}
		\centering
		\includegraphics[width=1\linewidth]{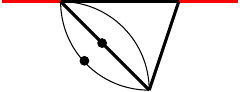}
		\caption*{$\M^{\text{NP1}}_{12}$}
	\end{minipage}
	\begin{minipage}{0.2\linewidth}
		\centering
		\includegraphics[width=1\linewidth]{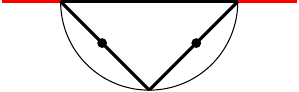}
		\caption*{$\M^{\text{NP1}}_{13}$}
	\end{minipage}
	\begin{minipage}{0.2\linewidth}
		\centering
		\includegraphics[width=1\linewidth]{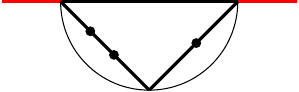}
		\caption*{$\M^{\text{NP1}}_{14}$}
	\end{minipage}
	\begin{minipage}{0.2\linewidth}
		\centering
		\includegraphics[width=1\linewidth]{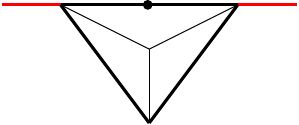}
		\caption*{$\M^{\text{NP1}}_{15}$}
	\end{minipage}
	\begin{minipage}{0.2\linewidth}
		\centering
		\includegraphics[width=1\linewidth]{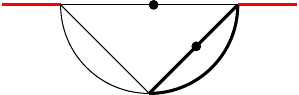}
		\caption*{$\M^{\text{NP1}}_{16}$}
	\end{minipage}
	\begin{minipage}{0.2\linewidth}
		\centering
		\includegraphics[width=1\linewidth]{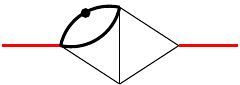}
		\caption*{$\M^{\text{NP1}}_{17}$}
	\end{minipage}
	\begin{minipage}{0.2\linewidth}
		\centering
		\includegraphics[width=1\linewidth]{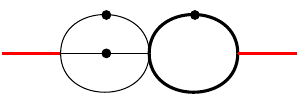}
		\caption*{$\M^{\text{NP1}}_{18}$}
	\end{minipage}
	\begin{minipage}{0.2\linewidth}
		\centering
		\includegraphics[width=1\linewidth]{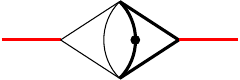}
		\caption*{$\M^{\text{NP1}}_{19}$}
	\end{minipage}
	\begin{minipage}{0.2\linewidth}
		\centering
		\includegraphics[width=1\linewidth]{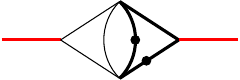}
		\caption*{$\M^{\text{NP1}}_{20}$}
	\end{minipage}
	\begin{minipage}{0.2\linewidth}
		\centering
		\includegraphics[width=1\linewidth]{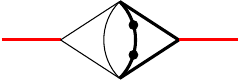}
		\caption*{$\M^{\text{NP1}}_{21}$}
	\end{minipage}
	\begin{minipage}{0.2\linewidth}
		\centering
		\includegraphics[width=1\linewidth]{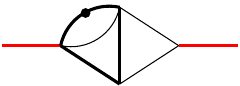}
		\caption*{$\M^{\text{NP1}}_{22}$}
	\end{minipage}
	\begin{minipage}{0.2\linewidth}
		\centering
		\includegraphics[width=1\linewidth]{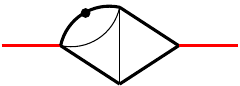}
		\caption*{$\M^{\text{NP1}}_{23}$}
	\end{minipage}
	\begin{minipage}{0.2\linewidth}
		\centering
		\includegraphics[width=1\linewidth]{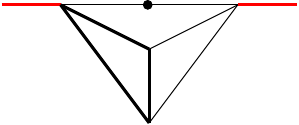}
		\caption*{$\M^{\text{NP1}}_{24}$}
	\end{minipage}
	\begin{minipage}{0.2\linewidth}
		\centering
		\includegraphics[width=1\linewidth]{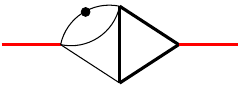}
		\caption*{$\M^{\text{NP1}}_{25}$}
	\end{minipage}
	\begin{minipage}{0.2\linewidth}
		\centering
		\includegraphics[width=1\linewidth]{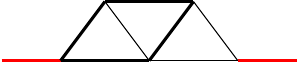}
		\caption*{$\M^{\text{NP1}}_{26}$}
	\end{minipage}
	\begin{minipage}{0.2\linewidth}
		\centering
		\includegraphics[width=1\linewidth]{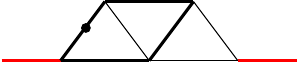}
		\caption*{$\M^{\text{NP1}}_{27}$}
	\end{minipage}
	\begin{minipage}{0.2\linewidth}
		\centering
		\includegraphics[width=1\linewidth]{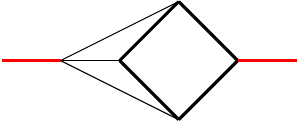}
		\caption*{$\M^{\text{NP1}}_{28}$}
	\end{minipage}
 	\begin{minipage}{0.2\linewidth}
		\centering
		\includegraphics[width=1\linewidth]{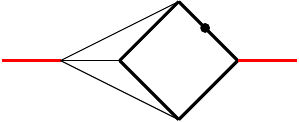}
		\caption*{$\M^{\text{NP1}}_{29}$}
	\end{minipage}
 	\begin{minipage}{0.2\linewidth}
		\centering
		\includegraphics[width=1\linewidth]{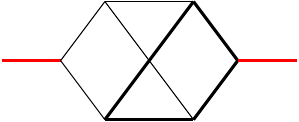}
		\caption*{$\M^{\text{NP1}}_{30}$}
	\end{minipage}
\caption{Master integrals in the NP1 topology. The thick black and red lines stand for the massive bottom quark and the Higgs boson, respectively. One black dot indicates one additional power of the corresponding propagator.}
\label{NP1_Topo}
\end{figure}

\begin{figure}[H]
	\centering
	\begin{minipage}{0.2\linewidth}
		\centering
		\includegraphics[width=1\linewidth]{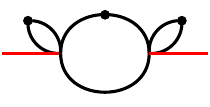}
		\caption*{$\M^{\text{P1}}_{1}$}
	\end{minipage}
	\begin{minipage}{0.2\linewidth}
		\centering
		\includegraphics[width=1\linewidth]{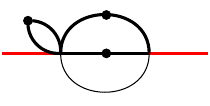}
		\caption*{$\M^{\text{P1}}_{2}$}
	\end{minipage}
	\begin{minipage}{0.2\linewidth}
		\centering
		\includegraphics[width=1\linewidth]{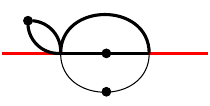}
		\caption*{$\M^{\text{P1}}_{3}$}
	\end{minipage}
	\centering
	\begin{minipage}{0.2\linewidth}
		\centering
		\includegraphics[width=1\linewidth]{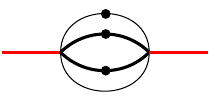}
		\caption*{$\M^{\text{P1}}_{4}$}
	\end{minipage}
	\begin{minipage}{0.2\linewidth}
		\centering
		\includegraphics[width=1\linewidth]{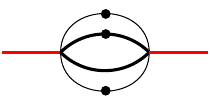}
		\caption*{$\M^{\text{P1}}_{5}$}
	\end{minipage}
	\begin{minipage}{0.2\linewidth}
		\centering
		\includegraphics[width=1\linewidth]{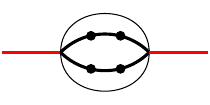}
		\caption*{$\M^{\text{P1}}_{6}$}
	\end{minipage}
	\begin{minipage}{0.2\linewidth}
		\centering
		\includegraphics[width=1\linewidth]{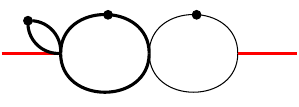}
		\caption*{$\M^{\text{P1}}_{7}$}
	\end{minipage}
	\begin{minipage}{0.2\linewidth}
		\centering
		\includegraphics[width=1\linewidth]{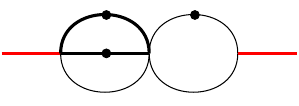}
		\caption*{$\M^{\text{P1}}_{8}$}
	\end{minipage}
	\begin{minipage}{0.2\linewidth}
		\centering
		\includegraphics[width=1\linewidth]{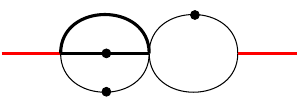}
		\caption*{$\M^{\text{P1}}_{9}$}
	\end{minipage}
	\begin{minipage}{0.2\linewidth}
		\centering
		\includegraphics[width=1\linewidth]{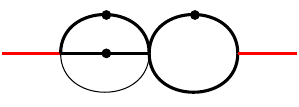}
		\caption*{$\M^{\text{P1}}_{10}$}
	\end{minipage}
	\begin{minipage}{0.2\linewidth}
		\centering
		\includegraphics[width=1\linewidth]{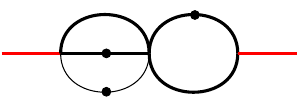}
		\caption*{$\M^{\text{P1}}_{11}$}
	\end{minipage}
	\begin{minipage}{0.2\linewidth}
		\centering
		\includegraphics[width=1\linewidth]{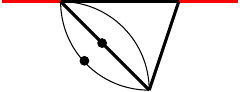}
		\caption*{$\M^{\text{P1}}_{12}$}
	\end{minipage}
	\begin{minipage}{0.2\linewidth}
		\centering
		\includegraphics[width=1\linewidth]{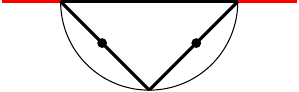}
		\caption*{$\M^{\text{P1}}_{13}$}
	\end{minipage}
	\begin{minipage}{0.2\linewidth}
		\centering
		\includegraphics[width=1\linewidth]{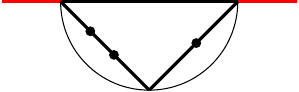}
		\caption*{$\M^{\text{P1}}_{14}$}
	\end{minipage}
	\begin{minipage}{0.2\linewidth}
		\centering
		\includegraphics[width=1\linewidth]{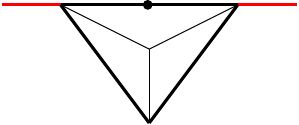}
		\caption*{$\M^{\text{P1}}_{15}$}
	\end{minipage}
	\begin{minipage}{0.2\linewidth}
		\centering
		\includegraphics[width=1\linewidth]{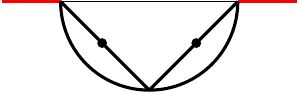}
		\caption*{$\M^{\text{P1}}_{16}$}
	\end{minipage}
	\begin{minipage}{0.2\linewidth}
		\centering
		\includegraphics[width=1\linewidth]{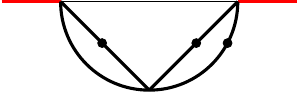}
		\caption*{$\M^{\text{P1}}_{17}$}
	\end{minipage}
	\begin{minipage}{0.2\linewidth}
		\centering
		\includegraphics[width=1\linewidth]{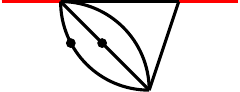}
		\caption*{$\M^{\text{P1}}_{18}$}
	\end{minipage}
	\begin{minipage}{0.2\linewidth}
		\centering
		\includegraphics[width=1\linewidth]{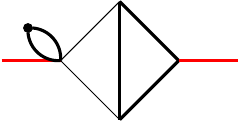}
		\caption*{$\M^{\text{P1}}_{19}$}
	\end{minipage}
	\begin{minipage}{0.2\linewidth}
		\centering
		\includegraphics[width=1\linewidth]{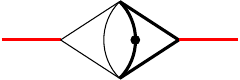}
		\caption*{$\M^{\text{P1}}_{20}$}
	\end{minipage}
	\begin{minipage}{0.2\linewidth}
		\centering
		\includegraphics[width=1\linewidth]{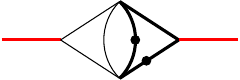}
		\caption*{$\M^{\text{P1}}_{21}$}
	\end{minipage}
	\begin{minipage}{0.2\linewidth}
		\centering
		\includegraphics[width=1\linewidth]{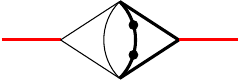}
		\caption*{$\M^{\text{P1}}_{22}$}
	\end{minipage}
	\begin{minipage}{0.2\linewidth}
		\centering
		\includegraphics[width=1\linewidth]{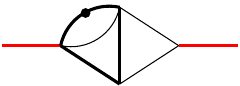}
		\caption*{$\M^{\text{P1}}_{23}$}
	\end{minipage}
	\begin{minipage}{0.2\linewidth}
		\centering
		\includegraphics[width=1\linewidth]{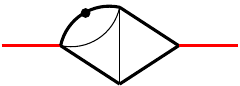}
		\caption*{$\M^{\text{P1}}_{24}$}
	\end{minipage}
	\begin{minipage}{0.2\linewidth}
		\centering
		\includegraphics[width=1\linewidth]{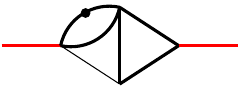}
		\caption*{$\M^{\text{P1}}_{25}$}
	\end{minipage}
	\begin{minipage}{0.2\linewidth}
		\centering
		\includegraphics[width=1\linewidth]{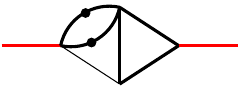}
		\caption*{$\M^{\text{P1}}_{26}$}
	\end{minipage}
	\begin{minipage}{0.2\linewidth}
		\centering
		\includegraphics[width=1\linewidth]{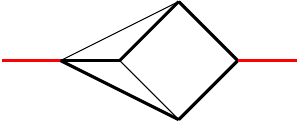}
		\caption*{$\M^{\text{P1}}_{27}$}
	\end{minipage}
	\begin{minipage}{0.2\linewidth}
		\centering
		\includegraphics[width=1\linewidth]{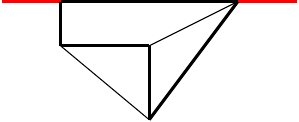}
		\caption*{$\M^{\text{P1}}_{28}$}
	\end{minipage}
 	\begin{minipage}{0.2\linewidth}
		\centering
		\includegraphics[width=1\linewidth]{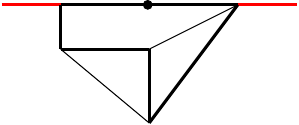}
		\caption*{$\M^{\text{P1}}_{29}$}
	\end{minipage}
 	\begin{minipage}{0.2\linewidth}
		\centering
		\includegraphics[width=1\linewidth]{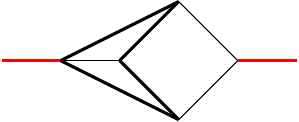}
		\caption*{$\M^{\text{P1}}_{30}$}
	\end{minipage}
  	\begin{minipage}{0.2\linewidth}
		\centering
		\includegraphics[width=1\linewidth]{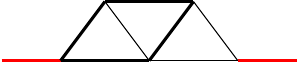}
		\caption*{$\M^{\text{P1}}_{31}$}
	\end{minipage}
  	\begin{minipage}{0.2\linewidth}
		\centering
		\includegraphics[width=1\linewidth]{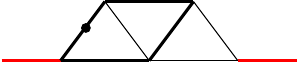}
		\caption*{$\M^{\text{P1}}_{32}$}
	\end{minipage}
\caption{Master integrals in the P1 topology. The thick black and red lines stand for the massive bottom quark and the Higgs boson, respectively. One black dot indicates one additional power of the corresponding propagator.}
\label{P1_Topo}
\end{figure}

The topology diagrams of the master integrals in the NP1 and P1 families 
are displayed in figure \ref{NP1_Topo} and figure \ref{P1_Topo}, respectively. 

\section{The canonical form of the three-loop kite integral family}
\label{sec:K3}

As mentioned in section~\ref{sec:MIs}, the master integral $\M^{\text{P1}}_{30}$, dubbed the three-loop kite integral, in the ${\rm P1}$ topology does \textit{not} contribute to the decay rate at $\mathcal{O}(\alpha_s^3)$. However, it may contribute at higher orders. 
In addition, the calculation of this integral and its sub-sectors is highly non-trivial and representative for multi-loop Feynman integrals. 
Specifically, it has the three-loop equal-mass banana integral family as its sub-sector. 
The geometry rooted in this three-loop banana integral family is a K3 surface, which is related to the symmetric product of the elliptic curve corresponding to the sunrise integral (two-loop equal-mass banana integral).
The higher sectors of the three-loop kite family have 7 master integrals and the related geometry  is a Riemann sphere\footnote{The K3 geometry enters only through the sub-sector dependence from the banana sector.}.
As a proof of concept, we derive the $\eps$-factorized differential equation for both the elliptic (or beyond elliptic) sub-sectors and higher sectors in this family. Our method is general and can be extended to other (two-scale) problems. Similar geometric structures have been discovered in
the two-loop kite integral~\cite{Sabry:1962rge,Remiddi:2016gno,Adams:2016xah}, in which only one master integral is present in the higher sector than the sunrise integrals. 
Recently, the three-loop corrections to the electron self-energy have been calculated in \cite{Duhr:2024bzt}, where the elliptic curve related to the sunrise integral plays an important role. 
However, there is no support for a cut on four massive particles and thus the three-loop banana integral does not appear as a sub-sector\footnote{Such a cut on four massive particles exists in the three-loop QED corrections to the photon propagator. 
There is a similar topology to the three-loop kite integral in our case. We thank Lorenzo Tancredi for pointing out this.}.

The three-loop kite family possesses 10 master integrals. The first three constitute the three-loop banana family\footnote{Since we impose a four-$b$ cut, the tadpole does not contribute. We hence suppress it from the start.}. The remaining 7 master integrals are in higher sectors. The structure of the differential equation for a pre-canonical basis reads
\begin{equation}
    \label{eq:P1prestruc}
    \frac{\rm d}{{\rm d} z}\begin{pmatrix}
{\color{red} \vec{I}_{\text{banana}}} \\
{\color{green} \vec{I}_{\text{higher}}} 
\end{pmatrix} =  
\begin{pmatrix}
\cellcolor{red} \mathbf{D}^{\text{banana}}_{3\times 3}(\eps, z) & \mathbf{0}_{3\times 7}\\
\cellcolor{orange} \mathbf{N}^{\text{banana}}_{7\times 3}(\eps, z) & \cellcolor{green} \mathbf{D}^{\text{higher}}_{7\times 7}(\eps, z) 
\end{pmatrix}\begin{pmatrix}
{\color{red} \vec{I}_{\text{banana}}} \\
{\color{green} \vec{I}_{\text{higher}}} 
\end{pmatrix},
\end{equation}
where we highlight the non-vanishing blocks in colors.
We first bring the diagonal blocks into $\eps$-form from lower to higher sectors, and then deal with non-diagonal blocks. We adopt the results in \cite{Pogel:2022ken} to bring the $3\times 3$ diagonal block for the banana integrals to the canonical form. Then we focus on the $7\times 7$ diagonal block, which involves the top integral, $\M^{\text{P1}}_{30}$. 
This is the most interesting part, and we will elaborate our method in detail, which can be useful for other Feynman integrals. In the end, we turn to the non-diagonal part, which describes the mixing with the banana family. 

Feynman integrals are characterized by their corresponding geometries, which can be detected from the maximal cuts (or the variety of graph polynomials).
A Riemann sphere corresponds to the functional space of MPLs.
An elliptic curve (torus) leads to the functional space of elliptic MPLs (eMPLs) and modular forms.
Higher-dimensional generalizations like Calabi-Yau $n$-folds or higher-genus generalizations like hyperelliptic curves indicate more complicated functional spaces.
Readers are referred to \cite{Bourjaily:2022bwx,Weinzierl:2022eaz} for reviews, and to \cite{Jiang:2023jmk,Gorges:2023zgv,Duhr:2024bzt} and references therein for recent examples. 

In our case, the three-loop banana sub-sector features a Calabi-Yau 2-folds, also called a K3 surface, while the higher sectors are related to a Riemann sphere except for the dependence on the banana sub-sector through mixing.  We propose an efficient and systematic way to find the $\eps$-factorized differential equation of the three-loop kite family, i.e., we apply the method developed to calculate the equal-mass banana integrals \cite{Pogel:2022ken,Pogel:2022yat,Pogel:2022vat} to the integrals characterized by a Riemann sphere\footnote{In the sense that the involved geometry is simpler, our method is a degeneration.} so that we can also use ``periods'' to construct the desired basis in higher sectors. In this unified language, we are able to give a criterion to choose a good pre-canonical basis and to explain the required rationalization in terms of the generalized ``mirror map''  in higher sectors.  

In the following, we briefly review existing results in the banana sub-sectors and then dive into the higher sectors. After that we deal with their mixing with the banana family.  We give the $\eps$-factorized differential equation for the entire family at the end.

\subsection{Banana sector}
The relevant pre-canonical master integrals in this sector are simply $I^{\rm P1}_{011100100}, I^{\rm P1}_{021100100}$ and $I^{\rm P1}_{031100100}$. The banana integrals are naturally set around the spacetime dimension $D=2$, which is related to the results around $D=4$ by the dimension-shift operator ${\bf D}^-$,
which lowers the dimension of spacetime by two through
\begin{equation}
    {\bf D}^- I^{\rm P1}_{\nu_1 \nu_2 \nu_3 \nu_4 \nu_5 \nu_6 \nu_7 \nu_8 \nu_9}\left( D \right) = 
 I^{\rm P1}_{\nu_1 \nu_2 \nu_3 \nu_4 \nu_5 \nu_6 \nu_7 \nu_8 \nu_9}\left( D-2 \right).    
\end{equation}
Then the canonical basis is given  by \cite{Pogel:2022yat}
\begin{equation}
    \begin{aligned}
       F_1 &= \frac{{\bf D}^- I^{\rm P1}_{011100100}(4-2\eps,z)}{\psi_1(z)},\\
       F_2 &= \frac{1}{\eps}\frac{1}{2\pi i} \frac{ {\rm d}}{{\rm d}\tau}F_1 +\left[G_2(z)+\frac{2(z-10)}{(z-4)(z-16)W(z)}\psi_1(z)\right]F_1,\\
       F_3 &= \frac{1}{\eps}\frac{1}{2\pi i} \frac{{\rm d}}{{\rm d}\tau} F_2 +\left[-2G_2(z)+\frac{2(z-10)}{(z-4)(z-16)W(z)}\psi_1(z)\right]F_2,\\
       &\hspace{2.6cm}+\left[\frac{3}{2}G_2^2(z)-\frac{(z+8)^2(z^2-8z+64)}{2z^2(z-4)^2(z-16)^2W^2(z)}\psi_1^2(z)\right]F_1,
    \end{aligned}
\end{equation}
with the mirror map
\begin{equation}
    \label{eq:Bananamirror}
	\tau = \frac{\psi_2(z)}{\psi_1(z)},\quad q=e^{2\pi i \tau},
\end{equation}
where $\psi_1$ and $\psi_2$ are periods of the K3 surface\footnote{It turns out that $\psi_1\propto \phi_1^2$ and $\psi_2 \propto \phi_1\phi_2$, where $\phi_1$ and $\phi_2$ are the two periods of an elliptic curve. In this sense, the K3 surface is called a symmetric product of the corresponding elliptic curve. } and are annihilated by the corresponding Picard-Fuchs operators. 
The Wronksian $W(z)$ reads
\begin{equation}
	W(z)  = \frac{-12}{z\sqrt{(z-4)(z-16)}}.
\end{equation}

The inverse relation between $\tau$ and $z$ is \cite{10.1215/kjm/1250518557}
\begin{equation}
    \label{eq:zq}
	z = -\left(\frac{\eta(\tau) \eta(3 \tau)}{\eta(2 \tau) \eta(6 \tau)}\right)^6,
\end{equation}
where $\eta(\tau)$ is the Dedekind eta function. 
The Jacobian from $z$ to the modular variable $\tau$ reads
\begin{equation}
    \label{eq:Jacobian}
	J(z)=\frac{1}{2\pi i}\frac{{\rm d}z}{{\rm d}\tau} = \frac{\psi_1(z)}{W(z)}.
\end{equation} 
The coefficient function $G_2(z)$ reads
\begin{equation}
    \label{eq:G3coeff}
    G_2(z)=(2 \pi i)^2 \int\limits_{i \infty}^\tau {\rm d} \tau_1 \int\limits_{i \infty}^{\tau_1} {\rm d} \tau_2 \frac{\tilde{z}(\tilde{z}-8)(\tilde{z}+8)^3}{864(4-\tilde{z})^{\frac{3}{2}}(16-\tilde{z})^{\frac{3}{2}}}\psi_1^6(\tilde{z}),
\end{equation}
where the integrand is viewed as a function of $\tau_2$ through $\tilde{z}(\tau_2)$ given by \eqref{eq:zq}. Then the canonical differential equation reads
\begin{equation}
    \frac{1}{2\pi i}\frac{{\rm d}}{{\rm d}\tau}\begin{pmatrix}
        F_1\\
        F_2\\
        F_3
    \end{pmatrix}=\eps\begin{pmatrix}
        -f_{2,a}-f_{2,b} & 1 & 0\\
        f_{4,b} & -f_{2,a}+2f_{2,b} & 1\\
        f_6 & f_{4,b} & -f_{2,a}-f_{2,b}
    \end{pmatrix}
    \begin{pmatrix}
        F_1\\
        F_2\\
        F_3
    \end{pmatrix}
    \label{eq:diffF123}
\end{equation}
with the above matrix elements being (meromorphic and/or quasi-Eichler) modular forms. Their explicit expressions, which can be found in \cite{Pogel:2022yat},  are not relevant for this work. 
In practice, one does not need to calculate \eqref{eq:G3coeff} and the above matrix elements in a closed form. But rather, one can perform expansion in $q$ via the mirror map. The evaluation of the iterated integrals in terms of $q$ is extremely fast. 

\subsection{Higher sectors}
Now we turn to the interesting part in this section, i.e., the higher sectors. Apart from the K3 geometry inherited from the sub-sector, it also has a (punctured) Riemann sphere inside, which is related to the function space of MPLs. 
It turns out that we can use the same method as in the banana family, i.e., constructing an ansatz using periods and the mirror map to change the variable, in order to derive the canonical form. In this context, the rationalization transformation in \eqref{eq:rationalization} and \eqref{eq:rationalizationinv} is dictated by the geometry. 

The complexity of Feynman integrals usually increases with the number of the propagators. 
Normally a pre-canonical basis can already transform the differential equation system into a lower block-triangular form. 
In our case, there are 7 master integrals apart from the three banana integrals after using \texttt{LiteRed} \cite{Lee:2013mka}. Since we are working around $D=4$, it is natural to raise the powers of some propagators. 
The following 7 pre-canonical master integrals are judiciously chosen:
\begin{equation}
    \begin{aligned}
        M_4 &= \eps^3(1-2\eps)  \,\,\,m_b^2I^{\rm P1}_{011201200}(4-2\eps,z),\\
        M_5 &= \eps^4(1-2\eps)  \,\,\,m_b^2I^{\rm P1}_{111101200}(4-2\eps,z),\\
        M_6 &= \eps^3(1-2\eps)  \,\,\,m_b^4I^{\rm P1}_{111201200}(4-2\eps,z),\\
        M_7 &= \eps^5(1-2\eps)  \,\,\,m_b^2I^{\rm P1}_{111111100}(4-2\eps,z),\\
        M_8 &= \eps^4(1-2\eps)^2\,m_b^2I^{\rm P1}_{011111110}(4-2\eps,z),\\
        M_9 &= \eps^4(1-2\eps)  \,\,\,m_b^4I^{\rm P1}_{021111110}(4-2\eps,z),\\
        M_{10} &= \eps^5(1-2\eps)  \,\,\,m_b^2I^{\rm P1}_{111110110}(4-2\eps,z).
    \end{aligned}
\end{equation}
Note that the master integrals $M_7, M_8, M_9$ are reducible, and belong to the sub-sectors of $M_{10}$, which is proportional to $\M^{\text{P1}}_{30}$.
The normalization factors have been fixed such that $M_i$ do not have divergences, which can be easily determined from the differential equation.
With this setup, the $7\times 7$ diagonal block  in (\ref{eq:P1prestruc}) in the higher sectors looks like
\begin{equation}
    \label{eq:triangular}
    \mathbf{D}^{\text{higher}}_{7\times 7}(\eps, x) = \begin{pmatrix}
        \cellcolor{green!10}\mathbf{d}^{(1)}_{1\times1} & \mathbf{0}_{1\times2} & \mathbf{0}_{1\times3} & \mathbf{0}_{1\times 1}\\
        \mathbf{n}^{(1)}_{2\times 1} & \cellcolor{green!30}\mathbf{d}^{(2)}_{2\times2} & \mathbf{0}_{2\times3} & \mathbf{0}_{2\times1}\\
        \mathbf{n}^{(2)}_{3\times 1} & \mathbf{n}^{(3)}_{3\times 2} & \cellcolor{green!50}\mathbf{d}^{(3)}_{3\times3} & \mathbf{0}_{3\times1}\\
        0 & \mathbf{n}^{(4)}_{1\times 2} & \mathbf{n}^{(5)}_{1\times 3} &\cellcolor{green}\mathbf{d}^{(4)}_{1\times 1}
    \end{pmatrix}.
\end{equation}

The strategy is always firstly to bring the upper diagonal blocks (corresponding to a lower sector) into the $\eps$-factorized form, and then to deal with the corresponding off-diagonal mixing in the same row\footnote{This mixing is within the $7\times 7$ block, not the mixing with the banana family.}. After that, we turn to lower blocks (higher sectors) and repeat the procedure until the whole matrix is in the $\eps$-factorized form. 

The first $1\times1$ block $\mathbf{d}^{(1)}_{1\times1}$  is trivial. 
The $\mathcal{O}(\eps^0)$ part is removed by solving a first-order homogeneous differential equation. This equation is nothing but the simplest Picard-Fuchs equation (a degree-one differential equation)  with a solution $\omega_1(z)=\sqrt{(z-4)/z}$.  Thus the desired master integral reads
\begin{equation}
    F_4 = \frac{M_4}{\omega_1(z)} = \sqrt{\frac{z}{z-4}}M_4.
\end{equation}

We elaborate in detail on the next $2\times 2$ block. 
The fact that $\mathbf{d}_{2\times2}^{(2)}$ is not degenerate indicates $(M_5, {\rm d} M_5/{\rm d}z)$  furnishes a basis. 
Then we can easily derive a second-order differential operator for $M_5$ as\footnote{Readers are referred to the book \cite{Weinzierl:2022eaz} for details on the Picard-Fuchs operators in the context of Feynman integrals.}
\begin{equation}
    \begin{aligned}
        \bigg[\overbrace{\frac{{\rm d}^2}{{\rm d} z^2}+\frac{3 \eps z+3 z-10}{(z-4) z}\frac{{\rm d}}{{\rm d} z}+ \frac{(1+2 \eps) (\eps z+2 \eps+z-2)}{(z-4) z^2}}^{\hat{L}^{(\eps)}_{2,(2)}(z)}\bigg]M_5 = \text{lower-sector integrals}. 
    \end{aligned}
\end{equation}
It suffices to consider the maximal cut in this sector (i.e., taking the involved 6 propagators on-shell) so that the lower-sector integrals on the right-hand side vanish. The operator $\hat{L}^{(\eps)}_{2,(2)}(z)$ is the Picard-Fuchs operator for this integral. It turns out that it factorizes linearly when $\eps=0$, which is a typical property for integrals related to MPLs\footnote{For integrals beyond MPLs, the corresponding Picard-Fuchs operators do not factorize linearly.},
\begin{equation}
    \label{eq:PF22}
    \hat{L}^{(\eps=0)}_{2,(2)}(z) = \left[\frac{\rm d}{{\rm d}z}+\frac{2(z-3)}{z(z-4)}\right]\left[\frac{\rm d}{{\rm d}z}+\frac{1}{z}\right].
\end{equation}
$\hat{L}^{(\eps=0)}_{2,(2)}(z)$ has the following two solutions:
\begin{equation}
    \label{eq:periods2}
    \begin{aligned}
        \omega_{2,1}(z) = \frac{1}{z},\quad \omega_{2,2}(z) = \frac{1}{2\pi i}\frac{1}{z}\log\frac{2+\sqrt{z(z-4)}-z}{2},
    \end{aligned}
\end{equation}
where we have chosen a normalization for later convenience. The solutions can be obtained by the method of variation of parameters. The first solution $\omega_{2,1}(z)$ corresponds to the rightmost factorized operator in (\ref{eq:PF22}) and is holomorphic on the Riemann sphere, especially around the infinite, which is the point of maximal unipotent monodromy (MUM). We will explain in more detail about the MUM point later on. The second solution $\omega_{2,2}(z)$  has logarithmic singularity and is related to the leftmost factorized operator. We call these two solutions ``periods'', although the related geometric object is a Riemann sphere, which has no non-trivial 1-cycle to define the period integral\footnote{Such ``periods'' also play an important role in  decomposing the connection matrix into semi-simple and unipotent parts in \cite{Gorges:2023zgv}, e.g., eq. (3.13) therein.}. Then we can formally define the ``mirror map'' by
\begin{equation}
    \label{eq:mirrirtop}
    \tau_2 \equiv  \frac{\omega_{2,2}(z)}{\omega_{2,1}(z)},\,\,  q_2 = e^{2\pi i \tau_2} = \frac{2+\sqrt{z(z-4)}-z}{2}.
\end{equation}
It is remarkable that the variable $q_2$, an analogue of $q$ in the elliptic case, is just the variable $w$ in eq. (\ref{eq:rationalizationinv}), which is introduced to rationalize the square root $\sqrt{z(z-4)}$. 
This procedure provides a geometric way to find the proper variable transformation for rationalization of square roots that appear in the differential equations.

Now we can write a similar ansatz for the canonical basis in this sector as in the elliptic case:
\begin{equation}
    \label{eq:block2ansatz}
    \begin{aligned}
        F_{5} = \frac{M_5}{\omega_{2,1}(z)},\quad F_6 = \frac{1}{\eps}\frac{1}{2\pi i}\frac{\rm d}{{\rm d}\tau_2} F_{5} - t_{11}(z)F_{5},
    \end{aligned}
\end{equation}
where $t_{11}(z)$ is a rotation coefficient depending on $z$. 
By requiring the differential equation is in $\eps$-factorized form, it is  determined to be 
\begin{equation}
    t_{11}(z) = -\sqrt{\frac{z-4}{z}}. 
\end{equation}
The mixing of this $2\times 2$ sector with $F_{4}$ is simple and can be cast in $\eps$-factorized form by refining $F_6$ as
\begin{equation}
    F_6 = \frac{1}{\eps}\frac{1}{2\pi i}\frac{\rm d}{{\rm d}\tau_2} F_{5} - t_{11}(z)F_{5} - \frac{3z}{2}F_{4}. 
\end{equation}

The most non-trivial part is the $3\times 3$ block $\mathbf{d}^{(3)}_{3\times 3}$. 
The Picard-Fuchs operators for a general $3\times 3$ differential matrix should have degree 3 and will factorize into linear operators when $\eps=0$ if the corresponding solution can be written in terms of MPLs. 
The Picard-Fuchs operator for $M_9$ is of degree 3. 
Setting $\eps=0$, we have 
\begin{equation}
    \label{eq:PFM9}
    \begin{aligned}
        \bigg[\overbrace{\frac{{\rm d}^3}{{\rm d} z^3}+\frac{8z-22}{z(z-4)}\frac{{\rm d}^2}{{\rm d} z^2}+\frac{14z-22}{z^2(z-4) }\frac{{\rm d}}{{\rm d} z}+ \frac{4z-2}{z^3(z-4) }}^{\hat{L}^{(\eps=0)}_{3,(3)}(z)}\bigg]M_9 = \text{lower-sector integrals}+\mathcal{O}(\eps). 
    \end{aligned}
\end{equation}
We find that this operator factorizes into three linear operators. 
In principle, one uses the information from this operator to construct a basis in this sector. 
This operator has three regular singularities: $\{0, 4,\infty\}$. 
Around these regular singularities, one can use the Frobenius method to construct three solutions in terms of series expansions \cite{agarwal2008ordinary}. 
The leading behavior of the three solutions around each point $z_i$ is of the form $(z-z_i)^{\rho_{i,j}}$, where $z_i\in \{0, 4,\infty\}$ and $j=1, 2, 3$.

When all the indices are equal, $\rho_{i, 1} = \rho_{i, 2} = \rho_{i, 3}$, one of the three solutions is holomorphic around such a point\footnote{Note that $\rho_{i,j}$ may not be an integer, but a rational number. Then holomorphicity is a feature of the series with $(z-z_i)^{\rho_{i,j}}$ stripped off.}, 
while the second and third solution develops single and double logarithmic structure, respectively.
In this context, the logarithmic structure is maximal, and the regular singularity $z_i$ is called a MUM point. 
The operator with a MUM point is nice, since one can use it to define the desired mirror map. 
In the case of \eqref{eq:PF22}, the MUM point is $z=\infty$, around which the two solutions are holomorphic and singly logarithmic, respectively. 
They are used to formally define the generalized ``mirror map'' \eqref{eq:mirrirtop}. If $\hat{L}^{(\eps=0)}_{3,(3)}(z)$ processes a MUM point, then there will be one more generalized ``period'', which is doubly logarithmic. This period can be used to define an additional so-called $Y$-invariant \cite{bogner2013algebraiccharacterizationdifferentialoperators}, which has appeared in the calculation of Calabi-Yau related Feynman integrals, like the banana integrals starting from four loops \cite{Pogel:2022ken, Pogel:2022vat} and the ice-cone integrals \cite{Duhr:2022dxb}.
A first application of the $Y$-invariant in elliptic Feynman integrals can be found in \cite{Jiang:2023jmk}.
However, $\hat{L}^{(\eps=0)}_{3,(3)}(z)$ does not have a MUM point, as shown by the following table.
\begin{table}[ht]
    \centering
    \label{tab:RiemannPsymbol}
    \begin{tabular}{|c|c|c|c|}
        \hline
        & \(z = 0\) & \(z = 4\) & \(z = \infty\) \\
        \hline
        \(\rho_1\) & \(-1 / 2\) & \(0\) & \(1\) \\
        \hline
        \(\rho_2\) & \(-1\) & \(1\) & \(2\) \\
        \hline
        \(\rho_3\) & \(-1\) & \(-1 / 2\) & \(2\) \\
        \hline
    \end{tabular}
\caption{The leading behaviors of the three solutions around regular singularities of $\hat{L}_{3,(3)}^{(\eps=0)}$.
}
\end{table}
Hence, we conclude that it is not a nice operator, 
and its related Feynman integrals are not a proper pre-canonical basis to start with. 
We believe that this criterion for choosing a nice operator and canonical basis can be  applicable to other cases. 

Following the same logic, the Picard-Fuchs operator for $M_8$ has three singularities: $z=0,4$ and $\infty$. 
The first two are regular singularities, but not MUM points and the last one is not even a regular singularity because this operator is in fact a reduced operator of degree-2 for ${\rm d}M_8/{\rm d}z$. 
Therefore, we do not choose $M_8$ as a good pre-canonical master integral either. 

The Picard-Fuchs operator for $M_7$ is exactly the same as $\hat{L}^{(\eps)}_{2,(2)}(z)$, which has proven to be a nice operator. Here we reuse the previous results and the first two basis integrals can be tentatively chosen,
\begin{equation}
    \label{eq:block3ansatzpre}
    \begin{aligned}
        F_{7} = \frac{M_7}{\omega_{2,1}(z)},\quad F_8 = \frac{1}{\eps}\frac{1}{2\pi i}\frac{\rm d}{{\rm d}\tau_2} F_{7} - t_{11}(z)F_{7},
    \end{aligned}
\end{equation}
following the same form as \eqref{eq:block2ansatz}. 
Since $\hat{L}^{(\eps)}_{2,(2)}(z)$ is of degree-2, 
$M_7$ and its derivatives do not have any dependence on the third basis in this block.
The candidates for the third basis include  $M_8$,  ${\rm d}M_8/{\rm d}z$, $M_9$.
It turns out that $M_8$ or ${\rm d}M_8/{\rm d}z$ are not good, since their $\eps$-factorized differential equations are not of Fuchsian form.
So we choose $M_9$ to proceed. 
Its derivative depends on $F_8$ at $\mathcal{O}(\eps^0)$ and this dependence can be removed by a rotation of the bases.
The same trick applies to the mixing of this sector with lower sectors.
We obtain the canonical basis as
\begin{equation}
    \label{eq:block3ansatz}
    \begin{aligned}
        F_{7} &= \frac{M_7}{\omega_{2,1}(z)},\\
        F_8 &= \frac{1}{\eps}\frac{1}{2\pi i}\frac{\rm d}{{\rm d}\tau_2} F_{7} - t_{11}(z)F_{7}+t_{11}(z)F_5,\\
        F_9 &= z M_9 + \frac{1}{2}t_{11}(z)F_8. 
    \end{aligned}
\end{equation}

At last, we come to the last $1\times 1$ block, which is the top sector integral $M_{10}$. We get rid of the $\mathcal{O}(\eps^0)$ terms in the differential equation of $M_{10}$ by solving a trivial equation, and the differential equation is converted to the canonical form with the definition of the basis
\begin{equation}
    \label{eq:block4ansatz}
    F_{10} = z M_{10}. 
\end{equation}

So far we have not taken the mixing with the banana sector into account. We turn to this part in the following. 

\subsection{Mixing}
\label{sect:mixingwbanana}
With the above setup, all the mixing of higher sectors to the banana sector has been set in canonical form except for the following two parts,
\begin{equation}
    \label{eq:mixingterms}
    \begin{aligned}
        \frac{\rm d}{{\rm d}z}F_4 &= -\frac{(z+2)\psi_1(z)}{12(z-4)\sqrt{z(z-4)}}F_1+\cdots , \\
        \frac{\rm d}{{\rm d}z}F_6 &= -\frac{z(z-22)\psi_1(z)+z(z^2-20z+64)\psi_1'(z)}{24(z-4)\sqrt{z(z-4)}}F_1+\cdots ,
    \end{aligned}
\end{equation}
where the dots denote the $\eps$-factorized terms and $\psi_1'(z)={\rm d}\psi_1(z)/{\rm d}z$. 
Then we can rotate $F_4$ and $F_6$ a bit with $F_1$ to get rid of the above non-$\eps$-factorized terms,
\begin{align}
    \label{eq:rotatedF4F6}
        F_4 &= \sqrt{\frac{z}{z-4}}M_4 + G_3(z)F_1,\,\\
        F_6 &= \frac{1}{\eps}\frac{1}{2\pi i}\frac{\rm d}{{\rm d}\tau_2} F_{5} - t_{11}(z)F_{5} - \frac{3z}{2}F_{4} + \left[\frac{\sqrt{z(z-4)}(z-16)\psi_1(z)}{24(z-4)}-8G_3(z)\right]F_1,\notag
\end{align}
with
\begin{equation}
    \begin{aligned}
        G_3(z) 
        =-2\pi i \int\limits_{i\infty}^\tau{\rm d}\tau_1\frac{(\tilde{z}+2)\sqrt{\tilde{z}(\tilde{z}-16)}}{144(\tilde{z}-4)}\psi_1^2(\tilde{z}),
    \end{aligned}
\end{equation}
where the integrand is a function of $\tau_1$ via $\tilde{z}(\tau_1)$.

\subsection{$\eps$-factorized form}
\label{sect:canonical}
We have now obtained the $\eps$-factorized differential equation for the three-loop kite family.
The result for the banana sector is shown in (\ref{eq:diffF123}),
and the result for the higher sectors is given by 
\begin{equation}
	\label{eq:P1eps}
	\begin{aligned}
		\frac{\rm d}{{\rm d}z}\begin{pmatrix}
			F_4 \\
			\ldots\\
			F_{10} 
		\end{pmatrix}  = \eps\, \mathbf{A}(z)\begin{pmatrix}
			F_4 \\
			\ldots\\
			F_{10} 
		\end{pmatrix} + \eps\, \mathbf{B}(z)\begin{pmatrix}
			F_1 \\
			F_2 \\
			F_{3} 
		\end{pmatrix} 
	\end{aligned},
\end{equation}
where
\begin{equation}
	\label{eq:P1epsA}
	\begin{aligned}
		\mathbf{A}(z) = \left(
		\begin{array}{ccccccc}
 \frac{1}{4-z} & 0 & 0 & 0 & 0 & 0 & 0 \\
 \frac{4}{r} & \frac{1}{z} & \frac{1}{2 r} & 0 & 0 & 0 & 0 \\
 -\frac{24}{z-4} & -\frac{12}{r} & -\frac{4 (z-1)}{(z-4) z} & 0 & 0 & 0 & 0 \\
 \frac{8 (3 z-4)}{(z-4) z} & -\frac{12}{r} & 0 & -\frac{4 (z-1)}{(z-4) z} & -\frac{12}{r} & 0 & 0 \\
 -\frac{8}{r} & 0 & -\frac{1}{2 r} & \frac{1}{2 r} & \frac{1}{z} & 0 & 0 \\
 \frac{4}{r} & -\frac{2}{z} & 0 & -\frac{1}{2 r} & -\frac{2}{z} & -\frac{1}{z} & 0 \\
 0 & \frac{2}{z} & 0 & 0 & \frac{2}{z} & 0 & -\frac{1}{z} \\
\end{array}
\right)
	\end{aligned}
\end{equation}
with $r$ defined in (\ref{eq:rationalization}). 
The non-vanishing matrix elements  of $\mathbf{B}$ read
\begin{equation}
	\label{eq:P1epsB}
	\begin{aligned}
		\mathbf{B}_{41}(z) &= \frac{(z+2)\psi_1(z)}{6\sqrt{z(z-4)^3}}-G_3(z)\left[\frac{1}{z-16}+\frac{G_2(z)}{J(z)}\right],\quad  \mathbf{B}_{42}(z) = \frac{G_3(z)}{J(z)},\\
		\mathbf{B}_{51}(z) &= -\frac{(z-16)\psi_1(z)}{12(z-4)}+\frac{4G_3(z)}{\sqrt{z(z-4)}}\\
		\mathbf{B}_{61}(z) &= \frac{(z^2-26z+16)\psi_1(z)}{3\sqrt{z(z-4)^3}}+G_3(z)\left[-\frac{8 ((z-48) z+128)}{(z-16) (z-4) z}+16\frac{G_2(z)}{J(z)}\right],\\
		\mathbf{B}_{62}(z) &= -16\frac{G_3(z)}{J(z)},\\
		\mathbf{B}_{71}(z) &= \frac{(z-16)(5z-4)\psi_1(z)}{12\sqrt{z(z-4)^3}}-\frac{8(3z-4)G_3(z)}{z(z-4)},\\
		\mathbf{B}_{91}(z) &= \frac{(z-16)(3z+4)\psi_1(z)}{48z(z-4)}-\frac{4G_3(z)}{\sqrt{z(z-4)}}.
	\end{aligned}
\end{equation}
The indices of the matrix $\mathbf{B}_{ij}$ run over $i=4,...,10$ and $j=1,2,3$.
The above matrix elements seem complicated, but they have simple expansions in terms of $q$.
The solutions of the differential equations in terms of Chen-iterated integrals and a comparison with the results obtained in the calculation of three-loop corrections to the photon propagator\footnote{We thank Lorenzo Tancredi for sharing Felix Forner's thesis \cite{Forner2024} in private when we were submitting this manuscript.} are left to future work. 

\bibliographystyle{JHEP}
\bibliography{reference}
\end{document}